\documentclass[12pt]{article}
\usepackage{graphicx}
\usepackage{geometry}
\usepackage{mathtools}
\usepackage{xcolor}
\usepackage{cancel}
\usepackage{amssymb}
\usepackage{comment}
\usepackage{mathrsfs}
\usepackage{soul}
\usepackage{breqn}

 \def\norm#1{\vert #1 \vert}
 \def\Norm#1{\Vert #1\Vert}

 \title{Parametric approximations of fast close encounters  
   of the planar three-body problem as arcs of
 a focus-focus dynamics}
 
\author{
        Massimiliano Guzzo\\
        {\small Dipartimento di Matematica ``Tullio Levi-Civita''}\\
        {\small Universit\`a degli Studi di Padova}\\
        {\small Padova 35122 (PD) Italia}}
\date{\today}

\begin{document}
\maketitle

\begin{abstract}
  A gravitational close encounter of a small body with a planet
  may produce a substantial change of its orbital parameters which can
  be studied using the circular restricted three-body problem. In this
  paper we provide parametric representations of the fast close encounters
  with the secondary body of the planar CRTBP as arcs of non-linear focus-focus dynamics. The result is the consequence of a  remarkable factorization of the Birkhoff normal forms of the Hamiltonian of the problem represented with the Levi-Civita regularization.  The
  parametrizations are computed using two different sequences of Birkhoff normalizations of given order $N$. For each value of $N$, the Birkhoff
normalizations and the parameters of the focus-focus dynamics are represented by polynomials whose coefficients can be computed iteratively with a computer algebra system; no quadratures, such as those needed to compute action-angle variables of resonant normal forms, are needed. We also provide some numerical demonstrations of the method for values of the mass parameter representative of the Sun-Earth and the Sun-Jupiter cases. 
\end{abstract} 

\section{Introduction}

\begin{figure}[!]
\center{

\includegraphics[height=8cm,angle=0]{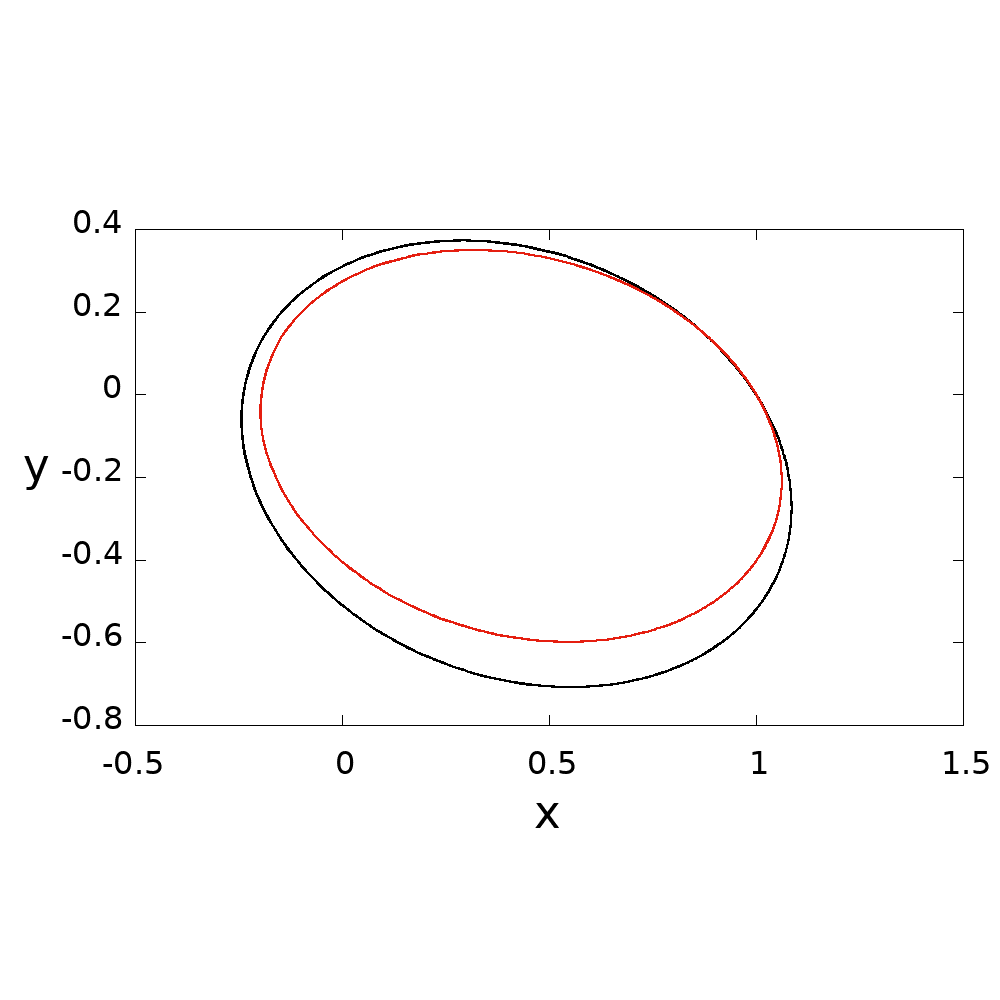}\includegraphics[height=8cm,angle=0]{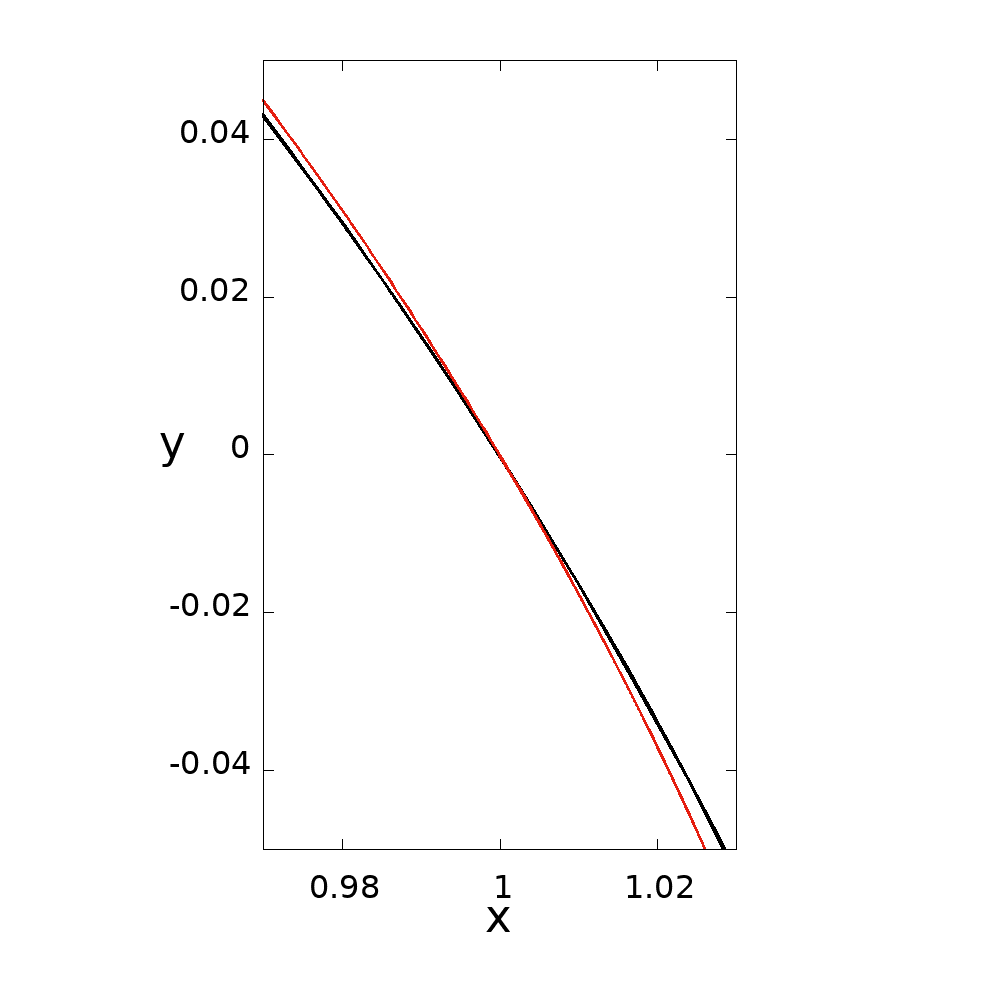}

\includegraphics[height=8cm,angle=0]{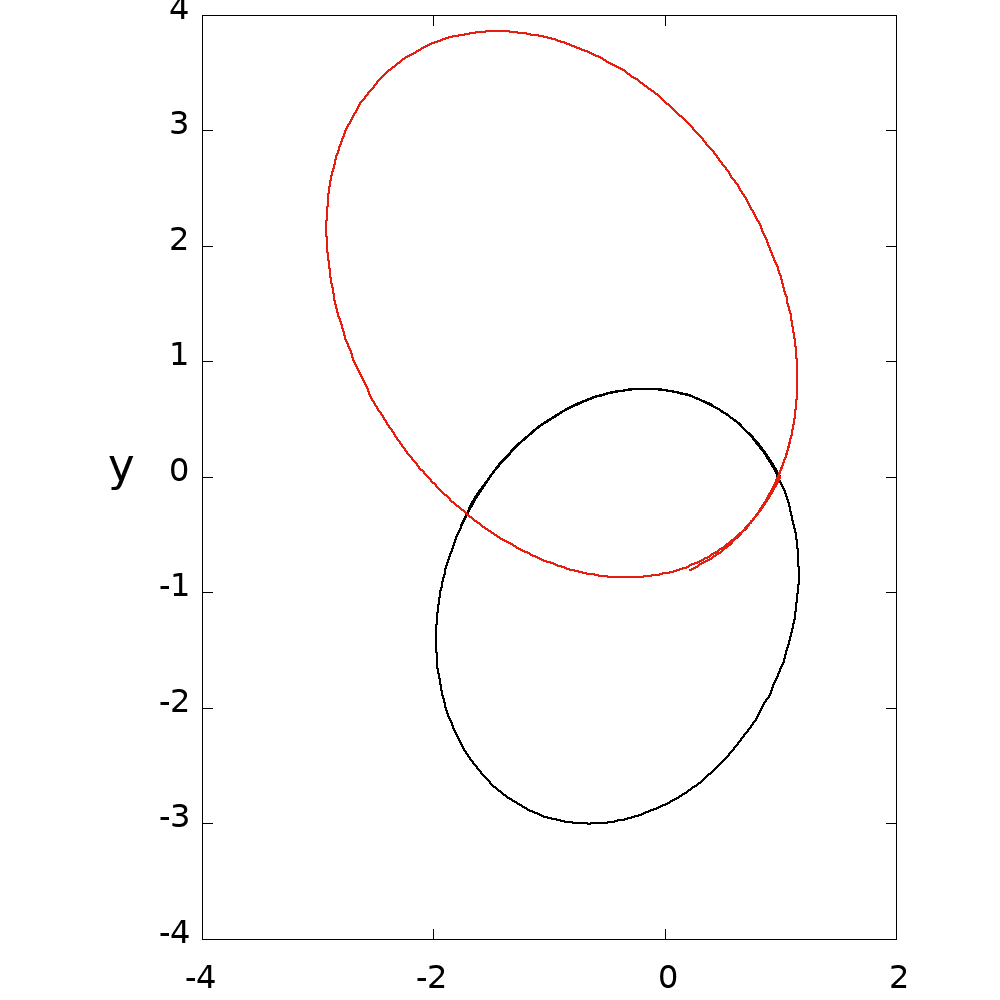}\includegraphics[height=8cm,angle=0]{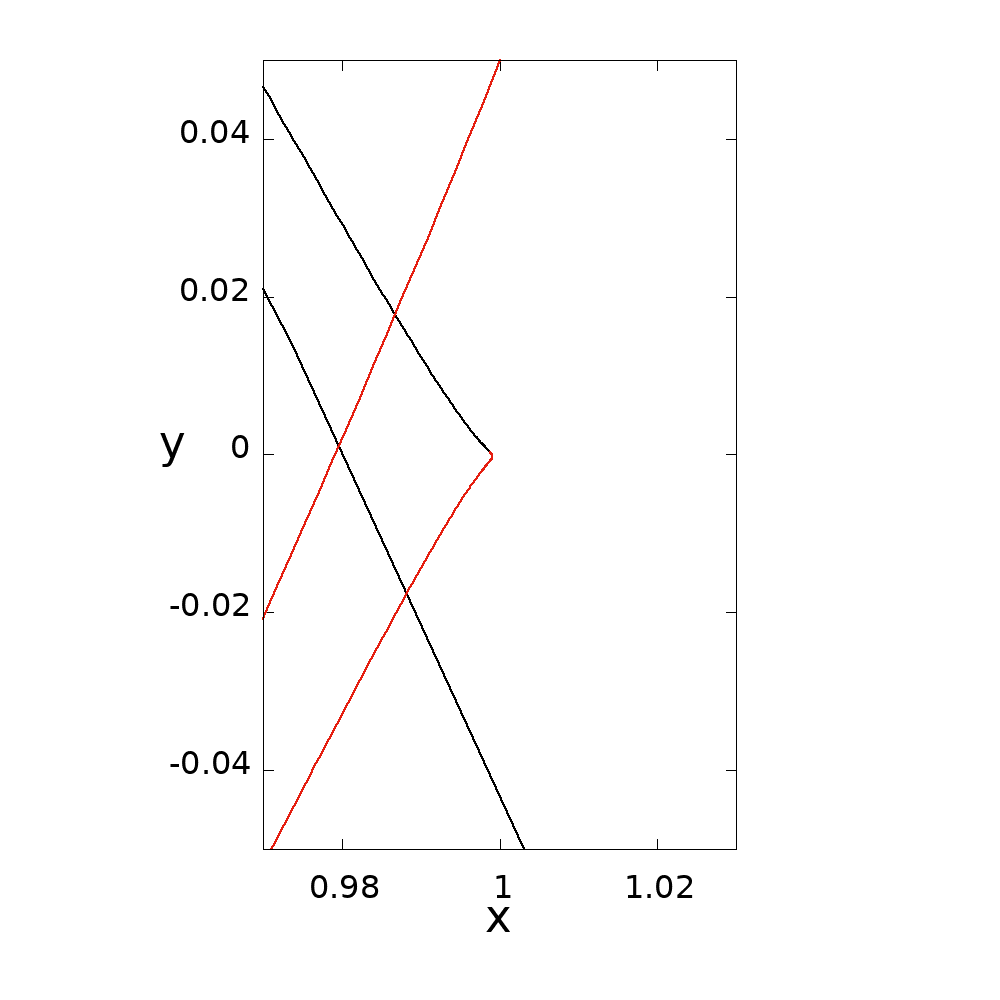}
  
}
\caption{Representation on the inertial orbital plane $Oxy$ of orbits with a close encounter at $t=0$: the black curves represent the orbits computed for positive times, the red curves represent the same orbits computed for negative times.
  On the top panels we represent an orbit computed
  for $\mu=3\ 10^{-6}$ (close to the Sun-Earth mass ratio)
  for $E=-1.35$ and initial conditions $u_1(0)=-10^{-2}$, $u_2(0)=10^{-2}$, $U_1(0)=-4 \ 10^{-6},U_2(0)= 0.016243781387232425 $; the close encounter determines
  a change of the semi-major axis of $\Delta a\sim 0.04349$ and $\Delta e\sim -0.07084$.
    On the bottom panels we represent an orbit computed
  for $\mu= 10^{-3}$ (close to the Sun-Jupiter mass ratio) for $E=-1.35$
  and  initial conditions $u_1(0)=10^{-2}$, $u_2(0)=2\ 10^{-2}$,
  $U_1(0)=-2\ 10^{-5},U_2(0)=0.092703055510000729$;  the close encounter determines
  a change of the semi-major axis of $\Delta a\sim -0.599$ and $\Delta e\sim -0.0708$.  The right panels represent a zoom close to the secondary body $P_2$.
  Details about the numerical computation of these orbits 
and additional analysis are given in Section \ref{sec:numerical}.}
\label{orbits}
\end{figure}

The problem of determining the effects of a gravitational
close encounter had been considered few years after the publication 
of Newton's {\it Philosophiae Naturalis Principia Mathematica}, in
an effort to better understand the apparitions of comets. 
Since comets are  visible from Earth when they are close to the Sun, their apparitions at different epochs
may be attributed to the same comet if they are linked by the same orbit. Using Newton's solutions of the 2-body problem Halley attributed the apparitions of 1531, 1607 and 1682  to the same
comet by conjecturing an heliocentric elliptic orbit, and predicted its subsequent return \cite{halley}; this prediction had been later refined   
by Clairaut by considering also the effects of planetary  perturbations \cite{clairaut}. A more puzzling situation manifested few years later, when the dramatic effects of close encounters with planet Jupiter had been
indicated as possible explanation for the appearance of a new comet
(Lexell's comet) and its subsequent disappearance\footnote{Lexell's comet, discovered in 1770, despite having an elliptic orbit of period of about  5.6 years, was not seen before as well as in the next 10 years (not either afterwards). A possible explanation was that the comet had not been seen before because of a close encounter with Jupiter in 1767, and it would be never be seen again because of a subsequent close encounter.}. The puzzling behaviour of Lexell'scomet, which had an explanation within the theory of the circular restricted
three-body problem, motivated the development of theories approximating the motion of Solar System bodies  transiting close to a planet by Laplace, Le Verrier and Tisserand \cite{Laplace,Leverrier,tisserand}. Since these pioneering papers the motion of a massless body $P$ having close encounters with a planet 
is conveniently approximated by an heliocentric motion of the 2-body problem defined by the primary body $P_1$ (the Sun), eventually modified by considering
the effects of planetary perturbations, as long as $P$ remains far from the secondary body $P_2$ (the planet). Instead, it is approximated by the motion of a
planetocentric 2-body problem defined by the secondary body $P_2$ during the short time of the close encounter. The switch between the different heliocentric
and planetocentric problems can produce a rapid and substantial alteration of
the heliocentric orbital parameters (see Figure \ref{orbits} for basic examples; more details will be given in Section \ref{sec:numerical}).
The analytic computation of the effects of close
encounters remains relevant for the modern astronomical applications which are related to  
the dynamics of comets (a remarkable example is provided by the dynamics of the comets of the Jupiter family, such as for comet 67/P Churyumov-Gerasimenko, target of the recent mission Rosetta), for the study of asteroids whose orbit represents a risk for potential Earth impacts, and for the modern space mission design where close encounters are used to modify the orbital elements of a spacecraft. There is therefore the problem of computing the change of the orbital parameters and
of representing parametrically the orbit during any individual close encounter.

There is a huge mathematical literature about collisions
(for example the ejection-collision orbits) and near
collision orbits in the restricted three-body problem
and related topics (for example, for near collision orbits of KAM type, such as the so-called punctured tori, see \cite{chencinerllibre,fejoz2002,zhao2015,GKZ}; for near collision orbits arising from studies
of Poincar\'e second species solutions see \cite{arenstorf,Perko,henrard,henon,marconiederman,bolotinmkkay2000,bolotin2006,FNS,FNS2009}; for computer assisted proofs see \cite{capinskietal2023}; and references therein). This paper is
about the representation of the arcs of solutions which intersect a small
neighbourhood of $P_2$, obtained from computations of series. 

Consider the planar circular restricted 
three-body problem defined by the motion of a body $P$ of infinitesimally 
small mass in the gravitation field of two massive bodies $P_1$
and $P_2$, the primary and secondary body respectively, 
which rotate uniformly around their common center of mass. In a rotating frame,  the Hamiltonian of the problem is:
\begin{equation}
h(x,y,p_x,p_y)= {p_x^2+p_y^2\over 2} + p_x y -p_y x 
-{1-\mu \over r_1}-{\mu\over r_2}  ,
\label{hambarycentric}
\end{equation}
where $r_1=\sqrt{(x+\mu)^2+y^2}$ and $r_2=\sqrt{(x-1+\mu)^2+y^2}$ denote 
the distances of $P$ from $P_1,P_2$ (as usual the units of mass,
length and time have been chosen so that the masses of $P_1$ and $P_2$ are  $1 - \mu$ and $\mu$ ($\mu \le 1/2$) respectively, their coordinates  are
$(x_1,0)=(-\mu,0)$, $(x_2,0)=(1-\mu,0)$ and their revolution period is $2\pi$). Let $\sigma$ be 
arbitrarily small; for any motion  
$(x(t),y(t))$ entering the ball centered at $P_2$ of radius $\sigma$  at time $t_0$ and leaving it at time $t_1$, the problem is to provide a parametric representation of the arc $(x(t),y(t),p_x(t),p_y(t))$ for $t\in [t_0,t_1]$.
For applications, a relevant point is
to formulate parametric representations which are valid
in neighborhoods defined by $\sigma$ as large as possible, 
for generic initial conditions (for example, without restricting the
analysis to symmetric orbits), and using methods which possibly
can be incrementally extended towards more realistic models of
the Solar System. Different rigorous approaches have been developed in the literature, reducing the problem to the computation of series. Within the methods
which exploit the regularizations at the secondary body $P_2$, 
we recall that a representation of close encounters for the planar problem had already been given in
the paper by Levi-Civita \cite{LC1906} (generalized to the spatial problem in \cite{CG21}), obtained from series representations of local solutions of the Hamilton-Jacobi equation
of the regularized Hamiltonian. A different representation  has been introduced by Henrard \cite{henrard} who defined Birkhoff normalizations of the Hamiltonian regularized at $P_2$. For representations using the non-regularized equations of motion we quote the methods of \cite{arenstorf} and \cite{Perko}, and subsequent developments.

Let us give more details about the method of \cite{henrard}, which applies to
the class of fast close encounters, i.e. the encounters occurring for values
 $E$ of the Hamilton function (\ref{hambarycentric})
satisfying\footnote{For comparison with
  paper \cite{henrard}, we recall that the value of the parameter $h$ appearing in Eqs. (10) and (11) of \cite{henrard} is related to $E$ by
  $E=h+\mu/2-\mu^2/2$. Also, the definition of the momenta $U_1,U_2$
  conjugate to the Levi-Civita variables $u_1,u_2$ introduced in \cite{henrard} is different from the definition given in Levi-Civita's paper \cite{LC1906},
  which we follow in this study. The two set of variables are related by
a canonical transformation.}:
\begin{equation}
  3+2E-4\mu+\mu^2:=\alpha^2 > 0  .
  \label{largeE}
\end{equation}
In \cite{henrard}, suitable resonant saddle-saddle Birkhoff normal forms of the Hamiltonian (\ref{hambarycentric}) represented with the Levi-Civita regularization had been considered.
At this regard, we introduce the
regularization of the Hamiltonian (\ref{hambarycentric})
following the method described in \cite{LC1906}: we 
first perform the phase-space translation:
\begin{equation}\label{eq:planetoXYZ}
X=x-x_{2}~, \quad Y=y~, \quad P_X=p_x~, \quad P_Y=p_y-x_{2}~~,
\end{equation}
and then we introduce the Levi-Civita variables $u=(u_1,u_2)$, 
extended to the conjugate momenta $U=(U_1,U_2)$ and to the fictitious time $\tau$,  
\begin{eqnarray}\label{eq:lc_var}
   & X = u_1^2-u_2^2~,   Y = 2u_1 u_2&~\\
  &P_X = \frac{U_1 u_1 - U_2 u_2}{2 |u|^2}~,
  P_Y = \frac{U_1 u_2 + U_2 u_1}{2 |u|^2}&~\\
    &dt = |u|^2 d\tau &
\end{eqnarray}
where $| {u}|^2 = u_1^2+u_2^2$. For any given value $E$ of the Hamiltonian (\ref{hambarycentric}),  the Levi-Civita Hamiltonian:
\begin{eqnarray}\label{eq:KofE}
\hskip -0.2 cm&{\cal K}_{E}( {u}, {U})= {1\over 8}\left (U_1+2\norm{ {u}}^2u_2\right )^2 + 
{1\over 8}\left (U_2-2\norm{ {u}}^2u_1\right )^2-{1\over 2}\norm{ {u}}^6-\mu &\cr
\hskip -0.2 cm&- | {u}|^2
\left(E + \frac{(1-\mu)^2}{2} \right) - (1-\mu) | {u}|^2
 \left[ \frac{1}{\sqrt{1 + 2 (u_1^2-u_2^2) + | {u}|^4}} +
    u_1^2 - u_2^2 \right]& ,
\end{eqnarray}
is a regularization of the planar circular restricted three-body problem at $P_2$,
in the sense that the solutions  $(u(\tau),U(\tau))$ of the Hamilton equations
of ${\cal K}_E$  satisfying
${\cal K}_{E}( {u(0)}, {U(0)})=0$ project, as long as
$\norm{u(\tau )}\ne 0$, on the solutions of the Hamilton
equations of (\ref{hambarycentric}), up to the
reparametrization of the time $t=t(\tau)$.

The Levi-Civita Hamiltonian (\ref{eq:KofE}) has
an equilibrium point at $(u,U)=(0,0,0,0)$ which,
as it was noticed in \cite{henrard}, is hyperbolic
for values $E$ satisfying (\ref{largeE}); in particular
the Jacobian of the Hamiltonian vector field of (\ref{eq:KofE})
computed at the equilibrium has eigenvalues $\pm \alpha/2$ of multiplicity 2.
This property has been exploited in the paper \cite{henrard} by
considering resonant hyperbolic Birkhoff normal forms (Bnf hereafter) of the regularized Hamiltonian. In \cite{henrard},
the Bnf have been introduced to discuss the application
of Hartman's theorem to the equilibrium point $(u,U)=(0,0,0,0)$;
therefore no high order Bnf have been constructed, as it
would be required by an high precision representation of the close
encounters.

We recall that Birkhoff normal forms offer an highly effective method to approximate the solutions which are close to the equilibrium points of
Hamiltonian systems. The method is useful also when the equilibrium is hyperbolic, since in this case it allows to  study the stable/unstable manifolds, as well as 
the transits close to the equilibrium; for example,
the method has been extensively used in the last decades to represent the solutions transiting close to the Lagrangian points $L_1,L_2$ (see, for example, \cite{Simo99,GJMS,JM99,masdemont05,CCP16,pucacco19,PG20,PG21,PG22,petersonetal2023,Rosalesetal2023}).
Depending on the resonant properties of Hamilton's equations linearized at the equilibrium, a suitable normal form Hamiltonian of arbitrary order $N$ is conjugate to the original one,
having the property that non-resonant monomials appear only at degrees
larger than $N$. For the case at hand, with a standard procedure
(which is recalled with all the details in Section \ref{chifirstBnf}), for any $N\geq 4$ we first define a canonical transformation:
$$
(u,U) \rightarrow (q,p)
$$
in a neighbourhood of $(u_*,U_*)=(0,0,0,0)$ (which is
a fixed point of the transformation) conjugating the Levi-Civita Hamiltonian
 (\ref{eq:KofE}) to 
\begin{equation}
K_E^N(q,p)=  -\mu + {\alpha\over 2}(q_1p_1+q_2p_2)+{\hat K}_4(q,p)+\ldots
+ {\hat K}_N(q,p)+{\cal R}_{N+2}(q,p)
\label{firstBNF}
\end{equation}
where the Taylor series of ${\cal R}_{N+2}$ in $q,p$ starts with
terms of degree at least $N+2$ and ${\hat K}_j$ are polynomials
of degree $j$ in $q,p$ (only
the polynomials with $j$ even number appear in the expansion)
containing monomials:
\begin{equation}
c^{j,N}_{m,n}q_1^{m_1}q_2^{m_2}p_1^{n_1}p_2^{n_2}
\label{monomial}
\end{equation}
with $m=(m_1,m_2),n=(n_1,n_2)$ satisfying:
\begin{equation}
n_1+n_2=m_1+m_2  .
\label{resnm}
\end{equation}
Relation (\ref{resnm}) is a direct consequence of the multiplicity of the characteristic
multipliers, which qualifies the origin as `resonant'. We will call 
`non resonant' the monomials (\ref{monomial}) which do not satisfy  relation
(\ref{resnm}); monomials which satisfy  relation (\ref{resnm}) will be called `integrable' if $n_1=m_1$ and $n_2=m_2$, `resonant' otherwise. 
The Hamilton function
which is obtained by neglecting in $K_E^N(q,p)$ the reminder ${\cal R}_{N+2}(q,p)$:
\begin{equation}
{\hat K}^N_E(q,p)=  -\mu + {\alpha \over 2}(q_1p_1+q_2p_2)+{\hat K}_4(q,p)+\ldots
+ {\hat K}^N(q,p) 
\label{firstbnfsenzar}
\end{equation}
will be called Bnf of order $N$. At variance with the full Hamiltonian
(\ref{firstBNF}), the Bnf (\ref{firstbnfsenzar})  is Poisson commuting
with the function $J(q,p)=q_1p_1+q_2p_2$ (this
a direct consequence of property (\ref{resnm}) of all the monomials
of the normal form), and its Hamilton equations are integrable by quadratures. For example, by introducing the canonical variables $I_j,\theta_j$ (defined for $I_j>0$; for $I_j<0$ obvious modifications are required): 
$$
q_j= \sqrt{I_j}e^{\theta_j}\ \ ,\ \ p_j=\sqrt{I_j}e^{-\theta_j}
$$
the integrable monomials are conjugate to functions depending only on the actions $I_j$,
while the resonant  monomials are conjugate to functions depending  on the actions $I_j$
and on multiples of $\theta_1-\theta_2$. As a consequence, the Hamiltonian
flow is integrable by quadratures (see Section \ref{sec:firstbnf} for
more details). Nevertheless,
working out the explicit solutions for any order $N$ is cumbersome,
because of the presence of the resonant terms (or equivalently, because of the
dependence on $\theta_1-\theta_2$).  The computation by quadratures may be particularly tricky to perform explicitly because the
resonant terms depend on different powers of the action variables $I_1,I_2$
and so the solution by quadratures
requires to represent the roots of an algebraic
polynomial in the action variables whose degree increases with the order $N$ of approximation.
This is a minor
problem if one is interested only in some low order approximation of
the flow, but it poses a serious challenge if one needs a parametric
representation of the flow which is valid for  arbitrary (or for
suitably large) order $N$. This problem does not exist for the Bnf
of Hamiltonian systems with an  
the equilibrium point which, in the linear approximation, is not resonant
(for example, this happens for equilibria which are linear elliptic
with Diophantine frequencies). In fact, in this case the Birkhoff
normal forms can be represented with a function depending only on
action variables, whose flow is easily represented.

\vskip 0.4 cm
\noindent
In this paper, we obtain a major improvement in the
computation of the solutions of the flow of the Bnf (\ref{firstbnfsenzar}).
Precisely, after proving a remarkable factorization property
of Hamiltonians  (\ref{firstbnfsenzar}) for all $N$, we are able to give
its Hamilton equations a different Hamiltonian representation,
where the origin $(u,U)=(0,0,0,0)$ is a non-resonant focus-focus
equilibrium point.
As a consequence, we are able to perform an
additional sequence of focus-focus Birkhoff normalizations 
conjugating the Hamiltonian to a normal form depending only on the action variables. Thus, we  obtain  a considerable simplification in the computation of close encounters, which is valid for small values of the mass parameter $\mu$. The factorization property that we prove for the Bnf is the following: for all $N$ the Birkhoff normal forms 
(\ref{firstbnfsenzar}) are divided by the polynomial
$J(q,p)=q_1p_1+q_2p_2$, i.e. they
are represented in the form:
\begin{equation}
\hat K^N_E(q,p)=  -
\mu+ (q_1p_1+q_2p_2) \left ( {\alpha \over 2} + k_2(q,p)+\ldots + k_{N-2}(q,p)\right )  ,
\label{bnffactor}
\end{equation}
where $k_j(q,p)$ are polynomials of degree $j$. Since both functions
$J(q,p)$ and $k(q,p)=  {\alpha \over 2} + k_2(q,p)+\ldots + k_{N-2}(q,p)$
are first integrals for the Birkhoff normal form  $\hat K^N_E$,
the solutions of Hamilton's equations with initial data satisfying
$$
k(q(0),p(0)):=\kappa \ \ \ \ , \ \ \ \ J(q(0),p(0)):=\eta \ \ ,
$$
with $\kappa \eta=\mu$, satisfy the Hamilton equations of:
\begin{equation}
{\cal H}(q,p)= \kappa (q_1p_1+q_2p_2)+\eta (k_2(q,p)+k_4(q,p)+\ldots + k_{N-2}(q,p) ) .
  \label{hametakappa0}
\end{equation}
Since $\eta$ is hereafter treated as a parameter, the terms $J(q,p)k_j(q,p)$ (which in the Hamiltonian (\ref{bnffactor}) are polynomials of degree $j+2$), are
transformed into polynomials $\eta k_j(q,p)$ which have degree
$j$; in particular $\eta k_2(q,p)$ has degree $2$ and modifies the
linearization of the Hamiltonian vector field at $(q,p)=(0,0,0,0)$.
Therefore, we have the opportunity to construct a different
Bnf for Hamiltonian (\ref{hametakappa0}) depending on this new
linearization.

For initial conditions satisfying $\hat K^N_E(q,p)=0$ we have $\eta\ne 0$, 
we therefore proceed by considering the case $\eta \ne 0$. 
By representing explicitly the term $k_2(q,p)$, we 
notice that Hamiltonian (\ref{hametakappa0}) has the
representation:
$$
{\cal H}(q,p) = \sum_{j\geq 1}^{N-2\over 2}h_{2j}(q,p)
$$
with quadratic part:
$$
h_2=\kappa (q_1p_1+q_2p_2)+\eta k_2(q,p) =
\Lambda (q_1p_1+q_2p_2)+\Omega (p_2 q_1-p_1 q_2)
$$
with $\Lambda=\kappa$ and $\Omega=\eta/(4\alpha)$,
while for $j\geq 2$ we have: $h_{2j}(q,p)= \eta k_{2j}(q,p)$. 
The origin  $(q,p)=(0,0,0,0)$ is an equilibrium point, with
characteristic exponents:
$$
\pm \Omega \pm i \Lambda ,
$$
so that the equilibrium is focus-focus with eigenvalues of multiplicity 1. 
 Therefore, we can proceed by defining a second sequence of Birkhoff normalizations of ${\cal H}$ using the focus-focus character of
 $(q,p)=(0,0,0,0)$. It is important to remark that the functions generating
 these Birkhoff normalizations have small divisors proportional to $\eta$, 
which may be very small. However, this does not
produce divergence of the norms since, by the factorization property,
the polynomial terms of order larger than 4 of the Bnf are proportional  
to $\eta$ as well, so that the generating functions of Birkhoff normalizations
are not singular at $\eta=0$. Moreover, the generating functions
depend in straightforward way on the parameters
$\Lambda,\Omega$ (see Section \ref{secondbnf} for details), so that the two additional parameters do not introduce any substantial complexity in the
symbolic computation of the normal forms. 

We therefore show that Hamiltonian (\ref{hametakappa0}) is conjugate
by a (complex) canonical transformation:
$$
(q,p) \mapsto (Q,P)
$$
defined in a neighbourhood of $(q,p)=(0,0,0,0)$ (which is
a fixed point of the transformation) to the focus-focus Bnf:
\begin{equation}
H^N_E(Q,P)=  (i\Omega -\Lambda)Q_1P_1-(i\Omega +\Lambda)Q_2P_2
+ {\hat h}(Q_1P_1,Q_2P_2)+{\cal R}_N(Q,P)
\label{secondBNF}
\end{equation}
where ${\hat h}(I_1,I_2)$ is a polynomial in $I_1,I_2$ containing
monomials of degree ranging between $2$ and $(N-2)/2$, and the
remainde ${\cal R}_N(Q,P)$ is a series which starts with terms of degree $N$. 
By neglecting the remainder we obtain the integrable Hamiltonian:
\begin{equation}
{\hat H}^N_E(Q,P)=  (i\Omega -\Lambda)Q_1P_1-(i\Omega +\Lambda)Q_2P_2
+ {\hat h}(Q_1P_1,Q_2P_2)
\label{secondbnfint}
\end{equation}
whose flow is represented explicitly by the same formulas, independently on $N$, since the functions $I_1=Q_1P_1,I_2=Q_2P_2$ are first integrals:
\begin{equation}
Q_j(\tau)=Q_j(0)e^{\hat \kappa_j \tau} \ \ ,\ \
P_j(t)=P_j(0)e^{-\hat\kappa_j \tau}\ \ ,\ \ j=1,2
\label{focusfocusdynamics0}
\end{equation}
where:
$$
\hat \kappa_j = {\partial \ \over \partial I_j} \left (
(i\Omega -\Lambda)I_1-(i\Omega +\Lambda)I_2
+ {\hat h}(I_1,I_2)\right )_{ \vert_{ I_1=I_1(0),I_2=I_2(0)}}  .
$$
For each order of approximation, all the Birkhoff
normalizations and the parameters of the focus-focus dynamics
(\ref{focusfocusdynamics0}) are represented by polynomials
whose coefficients can be computed iteratively with
a computer algebra system (see Sections \ref{chifirstBnf} and
\ref{sec:numerical}).

\vskip 0.4 cm
\noindent
    {\bf Remarks.} (I) Parametric representations of close encounters 
may be used  to implement a numerical integrator of the close
encounters by explicitly representing with a computer program the
flow of the Bnf $K^N_E(q,p)$.

\noindent
(II) The parametric representations of the solutions offered by the
Bnf are approximate, because they are obtained by neglecting
the remainders ${\cal R}_{N}(q,p)$, which are polynomials 
in the $(q,p)$ variables of order $N$. Therefore,
the distance of $(q,p)$ from the equilibrium point $(0,0,0,0)$ is the small parameter of the problem. We recall that the normal form variables
$q,p$ are obtained from the composition of
a linear transformation of the  Levi-Civita variables $(u,U)$ with the Birkhoff normalizations.  During a transit in a ball
of radius $\sigma$ centered at $P_2$, we have $\norm{u}^2< \sigma$
(which becomes $0$ at collision); but also $\norm{U}$ remains 
small (with limit value proportional to $\sqrt{\mu}$ at collision). 

\noindent
(III) As it is typical of normal form Hamiltonians, and
also of numerical integrators of given order, the remainder
${\cal R}_{N}(q,p)$ may depend on $N$ in a non-trivial case, so that
the increment of $N$ may produce an improvement only up to
a finite large value. For the numerical examples of Section \ref{sec:numerical}, obtained for the large order $N=30$, we see that we would still have the opportunity to reduce the error of the parametric representation by
considering also larger values of $N$. 
\vskip 0.4 cm

The paper is organized as follows: in Section 2 we describe
the properties of the first sequence of saddle-saddle Bnf; in Section 3 we prove that all the Birkhoff normal forms presented in Section 2 
are divided by $q_1p_1+q_2p_2$; in Section 4 we provide all the details needed to construct the second sequence of focus-focus Bnf, and the solution of
its Hamilton equations; in Section 5 we provide some numerical demonstrations of the method for values of the mass parameter representative of
the Sun-Earth and the Sun-Jupiter cases; in the Appendix 1 (Section 6)
we present the technical details of the construction of
the Birkhoff normal forms for 2-degrees of freedom Hamiltonian
systems; in the Appendix 2 (Section 7) we provide the generating functions
which define the Bnf of order $N=6$; Conclusions and Perspectives
are provided in Section 8.

\section{The resonant saddle-saddle Birkhoff normal form}\label{sec:firstbnf}

We expand the Levi-Civita Hamiltonian (\ref{eq:KofE}) as a Taylor
series of the variables $u,U$:
$$
{\cal K}_E = -\mu+{\cal K}_2(u,U;E)+{\cal K}_4(u,U)+
\sum_{j\geq 3} {\cal K}_{2j}(u)
$$
where:
$$
{\cal K}_2={1\over 8}(U_1^2+U_2^2) - {1\over 2}\left (3+2E-4\mu+\mu^2\right )(u_1^2+u_2^2)= {1\over 8}(U_1^2+U_2^2)-{\alpha^2\over 2}
(u_1^2+u_2^2) ,
$$
$$
{\cal K}_4= {1\over 2}(u_1^2+u_2^2)(U_1 u_2-U_2 u_1) ,
$$
and, for all $j\geq 3$, 
\begin{equation}
{\cal K}_{2j} =  - (1-\mu)\  T_{2j} \left [ | {u}|^2
 \left ( \frac{1}{\sqrt{1 + 2 (u_1^2-u_2^2) + | {u}|^4}}  \right )\right]  
\label{calK2j}
\end{equation}
 where $T_k f(u,U)$ denotes the Taylor term of degree $k$ of a function
 $f(u,U)$. For $E$ satisfying
 $$
 \alpha^2 = 3+2E-4\mu+\mu^2>0
 $$
 the origin is an hyperbolic equilibrium point with real eigenvalues
 $\pm {\alpha/ 2}$ of multiplicity 2. Therefore, we introduce hyperbolic variables by
means of the canonical transformation:
\begin{equation}
(q_1,q_2,p_1,p_2)=C^{-1}(u_1,u_2,U_1,U_2)
  \label{calC2}
  \end{equation}
where $C$ is  the real matrix:
$$
C=\left(
\begin{array}{cccc}
 \frac{1}{2\sqrt{\alpha}} & 0 & -\frac{1}{2\sqrt{\alpha}} & 0 \\
 0 & \frac{1}{2 \sqrt{\alpha}} & 0 & -\frac{1}{2 \sqrt{\alpha}} \\
 \sqrt{\alpha } & 0 & \sqrt{\alpha } & 0 \\
 0 & \sqrt{\alpha } & 0 & \sqrt{\alpha } \\
\end{array}
\right) .
$$
The transformation (\ref{calC2}) conjugates the Hamiltonian ${\cal K}_E$ to:
\begin{equation}
  K_E(q,p)=  -\mu + \sum_{j\geq 1}K_{2j}(q,p)
\label{KE}
\end{equation}
with:
$$
K_2 = {\alpha\over 2}(q_1p_1+q_2p_2)  
$$
and, for all $j$, we introduce the representation: 
$$
K_{2j} = \sum_{m,n \in {\Bbb N}^2: \norm{m}+\norm{n}=2j}c_{m,n}q^mp^n  ,
$$
where $m=(m_1,m_2),n=(n_1,n_2)$ are multi-indices and, 
 following standard notation for multi-indices, for
  any multi-index $\nu\in {\Bbb Z}^2$, 
  we denote $\norm{\nu}=\nu_1+\nu_2$.

  The Hamiltonian $K_E$ is given the  Birkhoff normal form
  by a canonical transformation ${\cal C}_N$, defined in a neighbourhood of $(q_*,p_*)=(0,0,0,0)$ (which is a fixed point of the transformation) conjugating
$K_E$ to:
\begin{equation}
K_E^N(q,p)=  -\mu + {\alpha\over 2}(q_1p_1+q_2p_2)+{\hat K}_4(q,p)+\ldots
+ {\hat K}_N(q,p)+{\cal R}_{N+2}(q,p)
\label{firstBNF1}
\end{equation}
where the Taylor series of ${\cal R}_{N+2}$ in $q,p$ starts with
terms of order at least $N+2$ and ${\hat K}_j$ are polynomials
of order $j$ in $q,p$ (only
the polynomials with $j$ even number appear in the expansion)
containing monomials:
\begin{equation}
c^{j,N}_{m,n}q_1^{m_1}q_2^{m_2}p_1^{n_1}p_2^{n_2}
\label{monomial1}
\end{equation}
with $m=(m_1,m_2),n=(n_1,n_2)$ satisfying:
\begin{equation}
n_1+n_2=m_1+m_2  .
\label{resnm1}
\end{equation}
The definition of the canonical transformation is defined following
a standard procedure, which we describe in detail in 
the Appendix 1 (Section \ref{chifirstBnf}).
\vskip 0.4 cm
\noindent
{\bf Remark.} The standard procedure
which is described in Section \ref{chifirstBnf} is
applied by setting the parameters $\lambda_1,\lambda_2=\alpha/2$,
so that the divisors  entering the definition
of the generating functions (\ref{eq:genfunc-a}):
$$
\lambda_1 (n_1-m_1)+\lambda_2 (n_2-m_2) = (\alpha/2)(n_1+n_2-m_1-m_2) ,
$$
are proportional to the integer $n_1+n_2-m_1-m_2$. As
a consequence, the resonant relation (\ref{resmonomialsgeneric}) becomes the
relation (\ref{resnm1}).
\vskip 0.4 cm
We consider the Hamiltonian which is obtained by neglecting the remainder ${\cal R}_{N+2}(q,p)$ in $K^N_E(q,p)$:
$$
{\hat K}^N_E(q,p)=  -\mu + {\alpha\over 2}(q_1p_1+q_2p_2)+{\hat K}_4(q,p)+\ldots
+ {\hat K}^N(q,p) ,
$$
which will be called Bnf of order $N$. For example, at order $N=6$ we have:
\begin{eqnarray}
&\hat K^6_E(q,p) =  -\mu +
     {\alpha\over 2}(q_1p_1+q_2p_2)-{1\over 4\alpha}(p_1 q_2-p_2 q_1)(q_1p_1+q_2p_2)&\cr
&    -{(p_1q_1+p_2q_2)\over
  16 \alpha^3}\Big [-5 (1-\mu)(q^2_1p^2_1+q^2_2p^2_2)
       +2 p_1p_2q_1q_2 (6-7\mu) &\cr
&       +(q^2_1p^2_2+q^2_2p^2_1)(4-3\mu)\Big ]  .&
\label{K6}
\end{eqnarray}
Higher order Bnf can be computed using modern computer algebra systems, with a
symbolic representation of the coefficients of all their 
monomials as a function of the parameters $\mu,E$. 

The Bnf of any order $N$ has the first integrals
${\hat K}^N_E(q,p)$ and $J(q,p)=q_1p_1+q_2p_2$ (this
a direct consequence of relation (\ref{resnm}) of all the monomials
of the normal form), and is integrable by
quadratures. In fact, by introducing the canonical variables $I_j,\theta_j$
(defined for $I_j>0$; for $I_j<0$ obvious modifications are required): 
$$
q_j= \sqrt{I_j}e^{\theta_j}\ \ ,\ \ p_j=\sqrt{I_j}e^{-\theta_j}
$$
the monomials of ${\hat K}^N_E$ are conjugate to:
$$
\sqrt{I_1}^{m_1+n_1}\sqrt{I_2}^{m_2+n_2}e^{(m_1-n_1) \theta_1}e^{(m_2-n_2) \theta_2}
= \sqrt{I_1}^{m_1+n_1}\sqrt{I_2}^{m_2+n_2}e^{(m_1-n_1) (\theta_1-\theta_2)}  ,
$$
where the last equality follows from relation (\ref{resnm}). With a further
change to the action-angle variables $\hat I_1,\hat I_2,\hat \theta_1,\hat
\theta_2$:
$$
\hat I_1=I_1\ \ ,\ \ \hat I_2=I_1+I_2\ \ ,\ \ \hat \theta_1=\theta_1-\theta_2
\ \ ,\ \ \hat \theta_2=\theta_2
$$
the  Bnf is finally conjugate to an Hamiltonian $h^N_E(\hat I_1,\hat I_2,\hat \theta_1)$ which is independent on $\hat \theta_2$, and therefore for any
value of the first integral $ \hat I_2=J(q,p)$ can be studied as a reduced 1-degree of freedom Hamiltonian system. Nevertheless, working out the explicit solutions for its Hamilton equations by quadratures,
which in particular requires to find the solutions
 of the equation:
\begin{equation}
h^N_E(\hat I_1,\hat I_2,\hat \theta_1) =0
\end{equation}
in the form $\hat I_1 := \hat I_1(\hat I_2,\hat \theta_1)$ for any
$N$, is cumbersome.

\section{A remarkable property of the Birkhoff normal forms}\label{property}

\subsection{A permutation symmetry}

We say that the polynomial $F(q,p)$ of the hyperbolic variables $q,p$:
\begin{equation}
F(q,p):= \sum_{\ell =1}^{L}\ \ \sum_{m,n:\ \norm{m}+\norm{n} =2\ell} c_{m,n}q^m p^n
\label{poliF}
\end{equation}
has the permutation symmetry if for each multi-index $m,n=m_1,m_2,n_1,n_2$ we have:
\begin{equation}
  c_{m,n}= (-1)^{\sigma_{m,n}+{\norm{m}+\norm{n}-2\over 2}} c_{\overline{m,n}} \ \  ,
  \label{permsym}
\end{equation}
where $\overline{{m,n}}=n_2,n_1,m_2,m_1$ and $\sigma_{m,n}=1$ is $m_1+n_1$ is odd, $\sigma_{m,n}=0$ otherwise.

\vskip 0.4 cm
We denote by $\Pi F$ the expansion obtained by summing all the
resonant terms of $F$:
$$
\Pi F :=  \sum_{\ell =1}^{L}\ \ \sum_{\norm{m}+\norm{n} =2\ell , 
  m_1+m_2=n_1+n_2} c_{m,n}q^m p^n  .
$$
We show that for any  polynomial $F(q,p)$ satisfying the permutation symmetry, $\Pi F$ is divisible by $q_1p_1+q_2p_2$. First, from the property (\ref{permsym}) we have:
\begin{itemize}
\item[(i)] $c_{m_1,m_2,m_2,m_1}=0$ for any $m_1,m_2$ with $m_1+m_2\geq 1$.
  In fact, for $n_1=m_2,n_2=m_1$, Eq.  (\ref{permsym}) becomes:
  $$
  c_{m_1,m_2,m_2,m_1} = (-1)^{\sigma_{m,n}}(-1)^{m_1+m_2-1}
  c_{m_1,m_2,m_2,m_1} .
  $$
 If $m_1+n_1=m_1+m_2$ is even, then $(-1)^{\sigma_{m,n}}=1$ and
 $(-1)^{m_1+m_2}=1$; if $m_1+n_1=m_1+m_2$ is odd, then $(-1)^{\sigma_{m,n}}=-1$ and $(-1)^{m_1+m_2}=-1$. In both cases we have
 $(-1)^{\sigma_{m,n}+{\norm{m}+\norm{n}-2\over 2}}=(-1)^{\sigma_{m,n}}(-1)^{m_1+m_2-1}=-1$, from which it follows $c_{m_1,m_2,m_2,m_1}=0$.
\item[(ii)] Consider any $m,n$ with $n_1,n_2\ne m_2,m_1$ and
  $m_1+m_2=n_1+n_2$. The two coefficients $c_{m,n},c_{\overline{m.n}}$
  satisfy:
  \begin{equation}
    c_{m,n}(-1)^{m_1}+c_{\overline{m,n}} (-1)^{n_2} =0 .
    \label{symmetriccoeff}
  \end{equation}
  In fact, since $m_1+m_2=n_1+n_2$ we rewrite Eq. (\ref{permsym})
  as:
  $$
  c_{m,n} = -(-1)^{\sigma_{m,n}}(-1)^{m_1+m_2}c_{\overline{m,n}}  .
  $$
  If $m_1+n_1,m_2+n_2$ are both even, on the one hand we have 
  $(-1)^{\sigma_{m,n}}=1$ so that:
  $$
  c_{m,n} = -(-1)^{m_1+m_2}c_{\overline{m,n}}  ,
  $$
  as well as:
  $$
  c_{m,n}(-1)^{m_1}+c_{\overline{m,n}}(-1)^{m_2}=0 .
  $$
  On the other hand, since $m_2+n_2$ is even, we have $(-1)^{m_2}= (-1)^{-n_2}=(-1)^{n_2}$,
  and so we obtain (\ref{symmetriccoeff}).
  If instead  $m_1+n_1,m_2+n_2$ are both odd, we have:
  $$
  c_{m,n} = (-1)^{m_1+m_2}c_{\overline{m,n}}  ,
  $$
  as well as:
  $$
  c_{m,n}(-1)^{m_1}-c_{\overline{m,n}}(-1)^{m_2}=0 .
  $$
  Since $m_2+n_2$ is odd, we have $(-1)^{m_2}= -(-1)^{-n_2}=-(-1)^{n_2}$,
  and so we obtain (\ref{symmetriccoeff}).
\end{itemize} 
From (i) and (ii) it follows that we can represent $\Pi F$ in
the form:
$$
\Pi F(q,p)= (q_1p_1+q_2p_2)  ( c + f_2(q,p)+\ldots + f_{N-2}(q,p))
$$
where $f_j(q,p)$ are polynomials of degree $j$ in the variables $q,p$.

In fact, in $\Pi F(q,p)$ we have all the monomials of $F$ 
corresponding to $m,n$ with $m_1+m_2=n_1+n_2$. If $m,n=\overline{m,n}$, from (i) we have $c_{m,n}=0$.
If $m,n\ne \overline{m,n}$ we consider the sum of the symmetric terms:
\begin{equation}
S_{m,n}=c_{m,n}q_1^{m_1}q_2^{m_2}p_1^{n_1}p_2^{n_2}+
c_{\overline{m,n}}q_1^{n_2}q_2^{n_1}p_1^{m_2}p_2^{m_1} ,
\label{twoterms}
\end{equation}
and prove that $S_{m,n}$ is divided by $q_1p_1+q_2p_2$.  Necessary
condition for a polynomial $S(q_1,q_2,p_1,p_2)$ to be divisible
by $(q_1p_1+q_2p_2)$ is that the rational function
${\cal S}=S(-q_2p_2/p_1,q_2,p_1,p_2)$ is identically zero. 
Since the ring of polynomials of four variables has unique factorization
(see \cite{Jacobson}, pages 153, 154), the condition is also sufficient. 
Let us therefore consider the rational function:
\begin{eqnarray*}
{\cal S}_{m,n} &=& S_{m,n}(-q_2p_2/p_1,q_2,p_1,p_2)\cr
&=& c_{m,n}(-1)^{m_1}q_2^{m_2+m_1}p_1^{n_1-m_1}p_2^{n_2+m_1}+
c_{\overline{m,n}}(-1)^{n_2}q_2^{n_1+n_2}p_1^{m_2-n_2}p_2^{m_1+n_2}  .
\end{eqnarray*}
Since the monomials of $\Pi F$ satisfy $m_1+m_2=n_1+n_2$, the powers
of $q_2,p_1,p_2$ are the same for both terms appearing in ${\cal S}_{m,n}$,
and therefore we have:
$$
{\cal S}_{m,n}=\left ( c_{m,n}(-1)^{m_1} + c_{\overline{m,n}}(-1)^{n_2} \right )
q_2^{m_2+m_1}p_1^{n_1-m_1}p_2^{n_2+m_1}  .
$$
By property (ii) the rational function ${\cal S}_{m,n}$ vanishes identically,
and therefore the polynomial $S_{m,n}$ in (\ref{twoterms}) is divided by $q_1p_1+q_2p_2$. Since $\Pi F$ is the sum of 
polynomials (\ref{twoterms}), it is divided by $q_1p_1+q_2p_2$ as well.

\subsection{The Levi-Civita Hamiltonian has the permutation symmetry}

In this Subsection we prove that any finite order truncation of the Levi-Civita Hamiltonian represented with the hyperbolic variables $q,p$, i.e. Hamiltonian
(\ref{KE}), has the permutation symmetry. First, one directly checks
that the terms:
$$
K_2= {\alpha\over 2}(q_1p_1+q_2p_2)
$$
$$
K_4 =\Big ( -\frac{  {q_1}{p_1}^2{p_2} }{8 \alpha} -\frac{ {q_1}{q_2}^2 {p_2} }{8 \alpha }\Big )
+\Big( -\frac{ {q_1}  {q_2} {p_1}^2 }{4 \alpha}-\frac{{q_2}^2 {p_1}  {p_2}}{4 \alpha }\Big )
+\Big (\frac{{q_2} {p_1}^3  }{8 \alpha }+\frac{  {q_2}^3 {p_1}}{8 \alpha} \Big )+
$$
$$
+\Big ( \frac{ {q_2} {p_1}{p_2}^2 }{8 \alpha }+\frac{  {q_1}^2{q_2} {p_1}}{8 \alpha }\Big )
+\Big (\frac{{q_1}^2 {p_1}  {p_2}}{4 \alpha }+\frac{ {q_1}  {q_2} {p_2}^2 }{4 \alpha}\Big )
+\Big (-\frac{  {q_1} {p_2}^3}{8 \alpha }
-\frac{{q_1}^3 {p_2}}{8 \alpha }\Big )
$$
have the permutation symmetry (the symmetric terms have been grouped together;
for all the terms of $K_4$ we have $\sigma_{m,n}=1$ and $\norm{m}+\norm{n}=4$,
so that (\ref{permsym}) becomes $c_{mn}=c_{\overline{m,n}}$).

To prove the
property for all the other terms $K_{2j}$ with $j\geq 3$,
we introduce the parametric function:
 $$
{\cal K}^{\geq 6}(u;\lambda)= 
-\sum_{\ell \geq 3}  (1-\mu)\  T_{2\ell} \left [ | {u}|^2
  \left ( \frac{1}{\sqrt{1 + 2\lambda (u_1^2-u_2^2) + \lambda^2| {u}|^4}}
\right)\right  ]
$$
and the Taylor expansion:
$$
{\cal K}^{\geq 6}(u;\lambda):=\sum_{j\geq 2}\lambda^j {\cal P}_j(u) ,\hfill
$$
where ${\cal P}_j(u)$ is a polynomial in the variables $u$ of degree $2j+2$; for $\lambda=1$ we have:
$$
 {\cal K}_E(u,U)=
-\mu +{\cal K}_2(u,U)+{\cal K}_4(u,U)+ {\cal K}^{\geq 6}(u;1) .
$$
The composition of the functions ${\cal K}^{\geq 6}$ and ${\cal P}_j$ with the canonical transformation:
$$
(u,U)=C(q,p)
$$
provides the functions $K^{\geq 6},P_j$:
$$
K^{\geq 6}(q,p;\lambda)= \sum_{j\geq 2}\lambda^j P_j(q,p)
$$
with $P_j(q,p)=K_{2j+2}(q,p)$ for all $j\geq 2$. 

For any $(q,p)$, by denoting with $(u,U):= C(q,p)$,
$(\tilde q_1,\tilde q_2,\tilde p_1,\tilde p_2):= (p_2,p_1,q_2,q_1)$
and $(\tilde u,\tilde U):= C(\tilde q,\tilde p)$, we have
$(\tilde u,\tilde U) = (-u_2,-u_1,U_2,U_1)$ as well as:
$$
{\cal K}^{\geq 6}(\tilde u;-\lambda)= {\cal K}^{\geq 6}
(u;\lambda) ,
$$
and consequently:
$$
\sum_{j\geq 2}\lambda^j {\cal P}_j(u) = \sum_{j\geq 2}\lambda^j
    (-1)^j {\cal P}_j(\tilde u) \ \ .
$$
By identifying the coefficients of the two Taylor expansions in $\lambda$
at both sides of the last equation, we obtain:
\begin{equation}
{\cal P}_j(u)=  (-1)^j {\cal P}_j(\tilde u)  ,
\label{eq-up1}
\end{equation}
as well as:
\begin{equation}
P_j(q,p)=  (-1)^j P_j(p_2,p_1,q_2,q_1)  .
\label{eq-qp1}
\end{equation}
From:
$$
P_j = \sum_{m,n:\norm{m}+\norm{n}=2j+2}c_{m,n}q^mp^n  ,
$$
by equating the coefficients of the same term $q^mp^n$
from both sides of equality (\ref{eq-qp1}) we obtain:
\begin{equation}
c_{m,n}= (-1)^j c_{\overline{m,n}}= (-1)^{\norm{m}+\norm{n}-2\over 2}
c_{\overline{m,n}} \ \ .
\label{permsym0}
\end{equation}
  Since the polynomials ${\cal P}_j$ depend on the $u$ only through $u_1^2,u_2^2$, in the corresponding polynomial $P_j(q,p)=K_{2j+2}(q,p)$ there are only monomials
  $c_{m,n}q^mp^n$ where $m_1+n_1,m_2+n_2$ are even numbers, and so
  $\sigma_{m,n}=0$. Therefore, eq. (\ref{permsym0}) can
  be rewritten as:
\begin{equation}
 c_{m,n}=   (-1)^{\sigma_{m,n}+{\norm{m}+\norm{n}-2\over 2}}c_{\overline{m,n}}  ,
\end{equation}
as it is required by (\ref{permsym}). Therefore we have proved that
all finite truncations of $K(q,p)$ have the permutation symmetry.

\subsection{The Birkhoff normal forms of the Levi-Civita Hamiltonian have the
  permutation symmetry}

In this Subsection prove that the Bnf (\ref{firstbnfsenzar}) of arbitrary order
$N$ have the permutation symmetry, and therefore are divisible
by $q_1p_1+q_2p_2$. 

Consider any couple of functions:
$$
F= \sum_{j\geq 1}\sum_{\norm{m}+\norm{n}= 2j}
c_{m,n}q^m p^n
$$
$$
\tilde F= \sum_{j\geq 1}\sum_{\norm{m}+\norm{n}= 2j}
\tilde c_{m,n}q^m p^n
$$
having the permutation symmetry, and the generating
function:
$$
\chi_{_{2i}} =\sum_{\norm{m}+\norm{n}= 2i,\norm{m} \ne \norm{n}}
{c_{m,n}\over {\alpha\over 2}(\norm{m}-\norm{n})}q^m p^n  
$$
which is constructed from the coefficients of the monomials
of degree $2i$ of the function $F$. We prove that $\{ \tilde F,\chi_{_{2i}}\}$  has the permutation symmetry.  It is sufficient to prove that, for any $j\geq 1$, by considering:
$$
\tilde F_{2j} =\sum_{\norm{\tilde m}+\norm{\tilde n}= 2j}
\tilde c_{\tilde m,\tilde n}q^{\tilde m}p^{\tilde n}  ,
$$
the function $\{\tilde F_{2j},\chi_{_{2i}}\}$ has the permutation symmetry.

We have:
\begin{eqnarray}
\{\tilde F_{2j},\chi_{_{2i}}\} & = 
\sum_{ \norm{m}+\norm{n}= 2i,\norm{m} \ne \norm{n}} 
\ \ \sum_{\norm{\tilde m}+\norm{\tilde n}= 2j}
   \ \ {\tilde c_{\tilde m,\tilde n} c_{m,n}\over
      {\alpha\over 2}(\norm{m}-\norm{n})} \times & \cr
&  \times  \Big [ (\tilde m_1 n_1-m_1\tilde n_1)
q_1^{m_1+\tilde m_1-1}q_2^{m_2+\tilde m_2}p_1^{n_1+\tilde n_1-1}
p_2^{n_2+\tilde n_2} &\cr 
&+  (\tilde m_2 n_2-m_2\tilde n_2)
q_1^{m_1+\tilde m_1}q_2^{m_2+\tilde m_2-1}p_1^{n_1+\tilde n_1}
p_2^{n_2+\tilde n_2 -1}\Big ]&\cr
   & := \sum_{ \norm{m}+\norm{n}= 2i,\norm{m} \ne \norm{n}} 
\ \ \sum_{\norm{\tilde m}+\norm{\tilde n}= 2j}
\Big [ {\cal P}_{m,n,\tilde m,\tilde n} +  {\cal Q}_{m,n,\tilde m,\tilde n} \Big ]   \ \ &
   \label{pbracket}
\end{eqnarray}
where:
$$
 {\cal P}_{m,n,\tilde m,\tilde n}:=  {\tilde c_{\tilde m,\tilde n} c_{m,n}\over
      {\alpha\over 2}(\norm{m}-\norm{n})} (\tilde m_1 n_1-m_1\tilde n_1)
q_1^{m_1+\tilde m_1-1}\ q_2^{m_2+\tilde m_2}p_1^{n_1+\tilde n_1-1}
p_2^{n_2+\tilde n_2}
$$
$$
  {\cal Q}_{m,n,\tilde m,\tilde n}:= {\tilde c_{\tilde m,\tilde n} c_{m,n}\over
      {\alpha\over 2}(\norm{m}-\norm{n})}(\tilde m_2 n_2-m_2\tilde n_2)
\ q_1^{m_1+\tilde m_1}q_2^{m_2+\tilde m_2-1}p_1^{n_1+\tilde n_1}
p_2^{n_2+\tilde n_2 -1}  .
$$
The double sum in (\ref{pbracket}) can be rearranged 
as the sum of couples of symmetric monomials 
${\cal P}_{m,n,\tilde m,\tilde n} :=
 \xi q_1^{M_1}q_2^{M_2}p_1^{N_1}p_2^{N_2}$ 
and ${\cal Q}_{\overline{m,n},\overline{\tilde m,\tilde n}}
:=  \eta q_1^{N_2}q_2^{N_1}p_1^{M_2}p_2^{M_1}$
satisfying:
\begin{equation}
\xi = (-1)^{\sigma_{M,N}+{\norm{M}+\norm{N}-2\over 2}}\eta ,
\end{equation}
which implies that $\{ \tilde F_j,\chi_{_{2i}}\}$ has the permutation symmetry.

In fact,  consider $m,n,\tilde m,\tilde n$ such that $m,n \ne \overline{m,n}$
and $\tilde m,\tilde n \ne \overline{\tilde m,\tilde n}$
(otherwise $\tilde c_{\tilde m,\tilde n} c_{m,n}=0$ and there is nothing more
to prove), and the term:
\begin{eqnarray}
{\cal P}_{m,n,\tilde m,\tilde n} &:=&  {\tilde c_{\tilde m,\tilde n} c_{m,n}\over
      {\alpha\over 2}(\norm{m}-\norm{n})} (\tilde m_1 n_1-m_1\tilde n_1)
q_1^{m_1+\tilde m_1-1}\ q_2^{m_2+\tilde m_2}p_1^{n_1+\tilde n_1-1}
p_2^{n_2+\tilde n_2}\cr
&=& \xi q_1^{M_1}q_2^{M_2}p_1^{N_1}p_2^{N_2}
\end{eqnarray}
where we denote:
\begin{equation}
M_1= m_1+\tilde m_1-1, M_2= m_2+\tilde m_2, N_1= n_1+\tilde n_1-1,
N_2= n_2+\tilde n_2 .
\label{monomialMN}
\end{equation}
and:
\begin{equation}
  \xi = {\tilde c_{\tilde m,\tilde n} c_{m,n}\over
    {\alpha\over 2}(\norm{m}-\norm{n})} (\tilde m_1 n_1-m_1\tilde n_1) .
  \end{equation}
We consider the symmetric term:
\begin{eqnarray}
{\cal Q}_{\overline{m,n},\overline{\tilde m,\tilde n}}&:=& {\tilde c_{\overline{\tilde m,\tilde n}} c_{\overline{m,n}}\over
      {\alpha\over 2}(\norm{n}-\norm{m})}(\tilde n_1 m_1-n_1\tilde m_1)
\ q_1^{n_2+\tilde n_2}q_2^{n_1+\tilde n_1-1}p_1^{m_2+\tilde m_2}
p_2^{m_1+\tilde m_1 -1}\cr
&=& \eta  q_1^{N_2}q_2^{N_1}p_1^{M_2}p_2^{M_1}
\end{eqnarray}
where $M_1,M_2,N_1,N_2$ have been defined in (\ref{monomialMN}), 
and:
$$
\eta ={\tilde c_{\overline{\tilde m,\tilde n}} c_{\overline{m,n}}\over
      {\alpha\over 2}(\norm{n}-\norm{m})}(\tilde n_1 m_1-n_1\tilde m_1)  .
$$
Since $F,\tilde F$ satisfy the
permutation symmetry we have:
$$
\xi = (-1)^{\sigma_{m,n}+\sigma_{\tilde m,\tilde n}+{\norm{m}+\norm{n}+\norm{\tilde m}+\norm{\tilde n}-4\over 2}}
    {\tilde c_{\overline {\tilde m,\tilde n}} c_{\overline{m,n}}\over
      {\alpha\over 2}(\norm{m}-\norm{n})}(\tilde m_1 n_1-m_1\tilde n_1)
$$
    $$
    =(-1)^{\sigma_{m,n}+\sigma_{\tilde m,\tilde n}+{\norm{M}+\norm{N}-2\over 2}}\eta  .
$$
    We have the two possibilities:
    \begin{itemize}
    \item[--] if $m_1+n_1$ and $\tilde m_1+\tilde n_1$ are both even or both odd we have $(-1)^{\sigma_{m,n}+\sigma_{\tilde m,\tilde n}}=1$; 
      $M_1+N_1=m_1+\tilde m_1+n_1+\tilde n_1-2$ is even and therefore: $(-1)^{\sigma_{m,n}+\sigma_{\tilde m,\tilde n}}=
      (-1)^{\sigma_{M,N}}$.
    \item[--] if  $m_1+n_1$ is even and $\tilde m_1+\tilde n_1$ is odd
(or $m_1+n_1$ is odd and $\tilde m_1+\tilde n_1$ is even), we have $(-1)^{\sigma_{m,n}+\sigma_{\tilde m,\tilde n}}=-1$;
      $M_1+N_1$ is odd therefore  $(-1)^{\sigma_{m,n}+\sigma_{\tilde m,\tilde n}}=
      (-1)^{\sigma_{M,N}}$.
          \end{itemize}
    Therefore we have:
$$
\xi   = (-1)^{\sigma_{M,N}+{\norm{M}+\norm{N}-2\over 2}}\eta  .
$$

Since the Birkhoff normal forms are constructed with Lie series
      of functions having the permutation symmetry (and the generating functions
      are constructed from the coefficients of functions having the permutation symmetry ) the BNF of any order $N$ has the permutation symmetry.

\section{A different Hamiltonian representation of the Birkhoff normal forms}\label{secondbnf}

Since the Bnf of any order $N$ has the permutation symmetry,
it is divisible by $J(q,p)=q_1p_1+q_2p_2$, i.e. we have: 
$$
\hat K^N_E(q,p)=  -\mu + {\alpha\over 2}(q_1p_1+q_2p_2)+{\hat K}_4(q,p)+\ldots
+ {\hat K}_N(q,p) =  -\mu+ (q_1p_1+q_2p_2) k(q,p)
$$
where:
$$
 k(q,p)={\alpha\over 2}  + k_2(q,p)+\ldots + k_{N-2}(q,p)
$$
 and $k_j$ are polynomials of order $j$ in the variables $q,p$. We remark that  all the monomials of $k(q,p)$ satisfy the resonant relation $m_1+m_2= n_1+n_2$ and moreover we have: $\{ k,  \hat K^N_E\} =  k \{ k ,J\} =0$. Therefore,
 both functions $J(q,p),k(q,p)$ are first integrals of the Bnf. 
 
Let us consider the Hamiltonian flow of the resonant Bnf of order $N$
with initial conditions satisfying $\hat K^N_E(q(0),p(0))=0$. 
The Hamilton's equations
of $\hat K^N_E$ have the form:
$$
\dot q_i= q_i\ k(q,p)+J(q,p)\ {\partial k\over \partial p_i}\ \ ,\ \ i=1,2
$$
$$
\dot p_i= -p_i\ k(q,p)-J(q,p)\ {\partial k\over \partial q_i}\ \ ,\ \ i=1,2
$$
and therefore the solutions with initial data satisfying
$$
k(q(0),p(0)):=\Lambda \ \ \ \ , \ \ \ \ J(q(0),p(0)):=\eta=4\alpha\Omega \ \ ,
$$
satisfy also the Hamilton equations of:
\begin{equation}
  {\cal H}(q,p)= \Lambda (q_1p_1+q_2p_2)+4\alpha\Omega (k_2(q,p)+k_4(q,p)+\ldots + k_{N-2}(q,p)) .
  \label{hametakappa}
\end{equation}
The condition $\hat K^N_E(q(0),p(0))=0$, equivalent to $4\alpha
\Lambda\Omega =\mu$,
excludes in particular the cases $\Lambda=0$ and $\Omega=0$. We therefore proceed by considering the case $\Omega \ne 0$ and we represent
Hamiltonian (\ref{hametakappa}) in the form:
$$
{\cal H}(q,p) = h_2(q,p)+ h_4(q,p)+\ldots +h_{N-2}(q,p) 
$$
with quadratic part (see Eq. (\ref{K6})):
$$
h_2:=  \Lambda (q_1p_1+q_2p_2) + \Omega (p_2q_1-p_1q_2)
$$
and $h_{2j}(q,p)=\eta k_{2j}(q,p)= 4\alpha
\Omega k_{2j}(q,p)$ for $j\geq 2$. 

 The origin  $(q,p)=(0,0,0,0)$ is an equilibrium point for the
 Hamiltonian flow of ${\cal H}$, with
characteristic exponents:
$$
\pm \Lambda \pm i\Omega
$$
so that the equilibrium is of focus-focus type. We therefore proceed by defining a second Birkhoff normalization of ${\cal H}$ using the focus-focus character of $(q,p)=(0,0,0,0)$.

\subsection{The focus-focus Birkhoff normal forms}

We first define the symplectic linear transformation:
\begin{eqnarray}
q_1 &=& {P_1-P_2\over \sqrt{2}} \ \ , \ \ p_1=  {Q_2-Q_1\over \sqrt{2}}\cr
q_2 &=& i{P_1+P_2\over \sqrt{2}} \ \ , \ \ p_2= i{Q_1+Q_2\over \sqrt{2}} ,
\label{qpQP}
\end{eqnarray}
which conjugates ${\cal H}$ to 
\begin{equation}
\hat {\cal H} :=\hat h_2(Q,P)+\hat h_4(Q,P)+\ldots +\hat h_{N-2}(Q,P) 
\label{hamQPnew}
\end{equation}
where:
$$
\hat h_2= (i\Omega -\Lambda)Q_1P_1-(i\Omega +\Lambda)Q_2P_2   
$$
and, for $j\geq 4$,  $\hat h_j$ is a polynomial of degree $j$  containing only monomials
\begin{equation}
c_{M,N}Q_1^{M_1}Q_2^{M_2}P_1^{N_1}P_2^{N_2} 
\end{equation}
with $M_1+M_2=N_1+N_2$. In fact, any monomial
$q_1^{m_1}q_2^{m_2}p_1^{n_1}p_2^{n_2}$  is conjugate by (\ref{qpQP}) to the polynomial:
$$
\left ( {P_1-P_2\over \sqrt{2}} \right )^{m_1}  \left ( i{P_1+P_2\over \sqrt{2}} \right )^{m_2} 
\left (  {Q_2-Q_1\over \sqrt{2}} \right  )^{n_1} \left (  i{Q_1+Q_2\over \sqrt{2}}\right  )^{n_2}  ,
$$
which is a sum of terms proportional to the monomials:
$$
P_1^{j_1}P_2^{m_1-j_1}P_1^{j_2}P_2^{m_2-j_2}
Q_1^{i_1}Q_2^{n_1-i_1}Q_1^{i_2}Q_2^{n_2-i_2}= Q_1^{M_1}Q_2^{M_2}P_1^{N_1}P_2^{N_2}
$$
with $M_1=i_1+i_2,M_2=n_1+n_2-i_1-i_2$, $N_1=j_1+j_2,N_2=m_1+m_2-j_1-j_2$;
from $m_1+m_2=n_1+n_2$ we have $M_1+M_2=N_1+N_2$.

The Hamiltonian $\hat {\cal H}$ is in the suitable form to apply
a sequence of Birkhoff normalizing canonical transformations following
the procedure which is described in detail in 
the Appendix (Section \ref{chifirstBnf}),  by setting the parameters
$\lambda_1=i\Omega -\Lambda,\lambda_2=-i\Omega -\Lambda$,
so that the divisors  entering the definition
of the generating functions (\ref{eq:genfunc-a})
are represented by:
\begin{equation}
\lambda_1 (N_1-M_1)+\lambda_2 (N_2-M_2) =
i \Omega (N_1-N_2 +M_2-M_1) -\Lambda (N_1+N_2-M_1-M_2) .
\label{smalldivnewQP}
\end{equation}
For $M,N$ satisfying $M_1+M_2=N_1+N_2$, the resonant relation (\ref{resmonomialsgeneric})
becomes:
$$
\lambda_1 (N_1-M_1)+\lambda_2 (N_2-M_2) = i \Omega (N_1-N_2 +M_2-M_1)=0
$$
which is satisfied if and only if $N_1=M_1$ and $N_2=M_2$.



Therefore, we have the opportunity to construct a non-resonant Birkhoff
normalization, i.e. a finite sequence of Birkhoff elementary canonical
transformations conjugating Hamiltonian (\ref{hamQPnew}) to
an Hamiltonian of the form:
\begin{equation}
H^N_E(Q,P)=  (i\Omega -\Lambda)Q_1P_1-(i\Omega +\Lambda)Q_2P_2
+ {\hat h}(Q_1P_1,Q_2P_2)+{\cal R}_N(Q,P)
\label{secondBNF}
\end{equation}
where ${\hat h}(I_1,I_2)$ is a polynomial in $I_1=Q_1P_1$,
$I_2=Q_2P_2$ containing
monomials of degree ranging between $2$ and $(N-2)/2$, and ${\cal R}_N(Q,P)$ has Taylor series starting at order $N$.

Therefore, the Bnf which is obtained by 
neglecting in (\ref{secondBNF}) the terms of order larger than $N-2$, is integrable since it contains only monomials satisfying $M_1=N_1$ and
$M_2=N_2$, i.e. it  depends on the variables $Q,P$ only through the actions $I_1=Q_1P_1$, $I_2=Q_2P_2$.

\vskip 0.4 cm
\noindent
{\bf Remark.} Even if the small divisors (\ref{smalldivnewQP})
are proportional to $\Omega$, which may be very small, the generating
function (\ref{eq:genfunc-a}) $\chi_{4}$ is defined
from the polynomial $\hat h_4$ which is the sum of monomials
proportional to $\Omega$ as well. Therefore
$\chi_4$ is independent of $\Omega$ and the Lie series: 
  $$
  \sum_{j\geq 1}{1\over j!}{\cal L}^j_\chi {\hat h}_2 =
    \sum_{j\geq 1}{1\over j!}{\cal L}^{j-1}_{\chi_4}(\Pi_0 \hat h_4 -\hat h_4) .
  $$
  $$
    \sum_{j\geq 0}{1\over j!}{\cal L}^j_{\chi_4} {\hat h}_{2\ell} ,\ \ \ \  \ell\geq 2
  $$
    are the sum of monomials proportional to $\Omega$. The argument
    is repeated at all orders. 
Therefore, at any step of the sequence of Birkhoff normalizations
we have generating functions which are independent on
$\Omega,\Lambda$, and the Lie series produced by the normalization 
depend on these parameter
in straightforward way.  This means that we do not need to introduce
  $\Lambda,\Omega$ as additional parameters (beyond a straightforward multiplicative factor) to compute symbolically the Birkhoff normal forms of Hamiltonian $\hat {\cal H}$.

\subsection{The Hamiltonian flow of the focus-focus Birkhoff normal form}

By neglecting the remainder from Hamiltonian (\ref{secondBNF}) we obtain the integrable Hamiltonian:
\begin{equation}
\hat H^N_E(Q,P)=  (i\Omega-\Lambda )Q_1P_1-(i\Omega+\Lambda )Q_2P_2
+ {\hat h}(Q_1P_1,Q_2P_2) 
\label{secondbnfint}
\end{equation}
which in particular is Poisson  commuting with $I_1=Q_1P_1$,
$I_2=Q_2P_2$. Since $I_1,I_2$ are first integrals, the Hamiltonian flow of (\ref{secondbnfint}) is represented explicitly by the formulas:
\begin{equation}
Q_j(\tau)=Q_j(0)e^{\hat \kappa_j \tau} \ \ ,\ \
P_j(t)=P_j(0)e^{-\hat\kappa_j \tau}\ \ ,\ \ j=1,2
\label{focusfocusdynamics}
\end{equation}
where:
$$
\hat \kappa_j = {\partial \ \over \partial I_j} \left (
(i\Omega -\Lambda)I_1-(i\Omega +\Lambda)I_2
+ {\hat h}(I_1,I_2)\right )_{ \vert_{ I_1=I_1(0),I_2=I_2(0)}}  .
$$
The representation of the solutions is now complete. For example,
for $N=8$, we have:
$$
\hat H^8_E(Q,P)=  (i\Omega -\Lambda)Q_1P_1-(i\Omega +\Lambda)Q_2P_2+
$$
$$
+{\eta \over 32\alpha^3}
\Big ((P_1^2Q_1^2+P_2^2Q_2^2) (3-\mu)+4Q_1P_1Q_2P_2 (1-2\mu )\Big )+
$$
$$
     {i \eta \over 64 \alpha^5}\left [ (5-3\mu)(P_2^3Q_2^3-P_1^3Q_1^3)
       -{3 \over 4} (95-178\mu+75 \mu^2)(P_1^2Q_1^2P_2Q_2-P_1Q_1P_2^2Q_2^2)\right ]  .
$$
We remark that, for each order of approximation, the Birkhoff
normalizations and the parameters of the focus-focus dynamics
(\ref{focusfocusdynamics}) are represented by polynomials
whose coefficients can be computed iteratively by a computer algebra system
(see Section \ref{appendix2}, Appendix 2, for the generating functions
of the Bnf of order $N=6$).

\section{Numerical demonstrations}\label{sec:numerical}

In this Section we provide some numerical demonstrations
of the application of the Birkhoff normalizations
described in the previous Sections. The goal is to show the
effectiveness of the representations
provided in this paper for some realistic choices
of the parameters $\mu,E$. A
systematic investigation of the effectiveness of the method 
for a wider range of applications is left to a further paper.

       \begin{figure}[!t]
\center{

\includegraphics[height=8cm,angle=0]{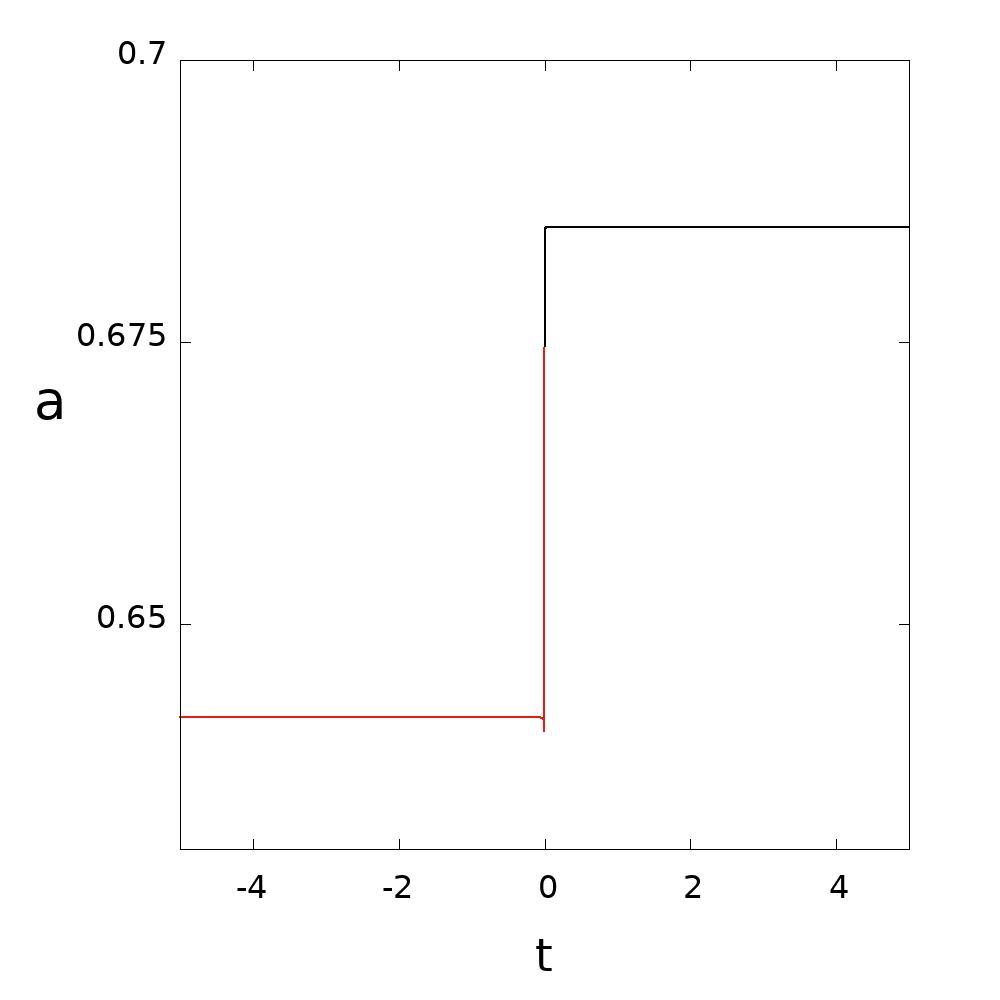}\includegraphics[height=8cm,angle=0]{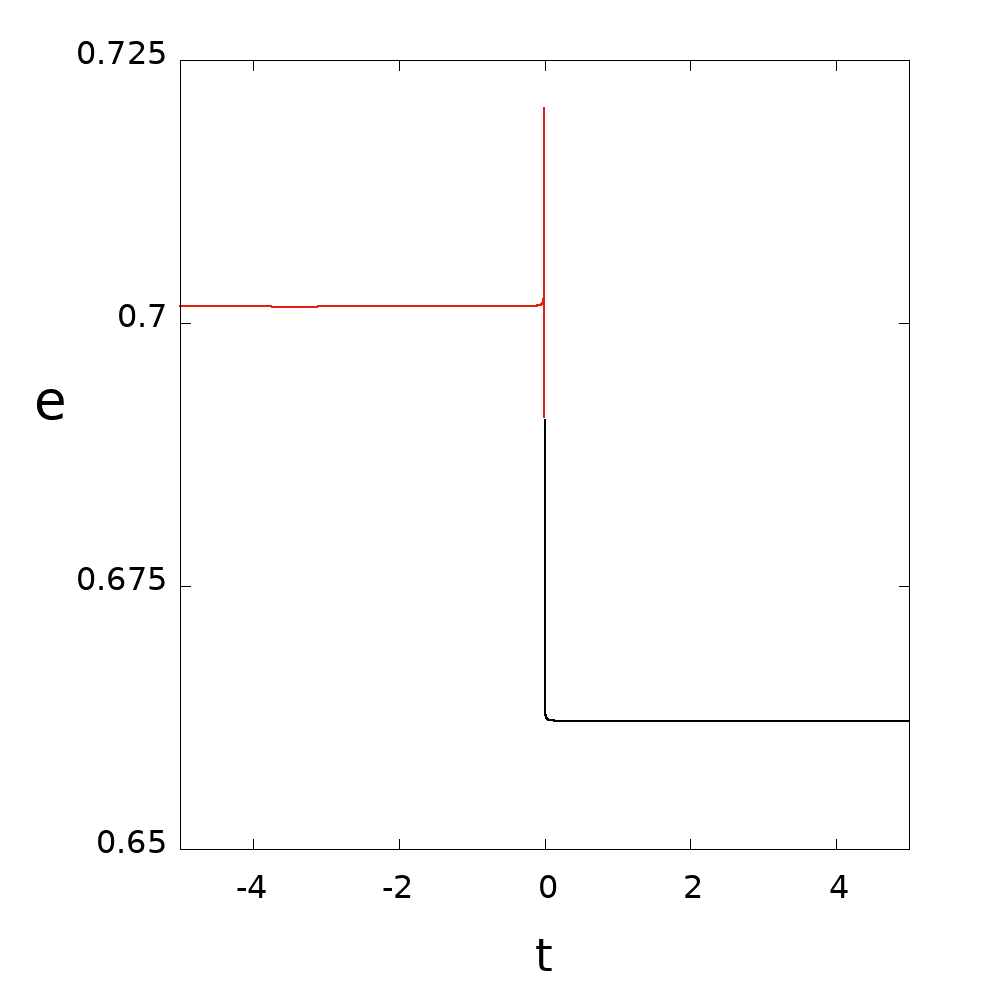}

  \includegraphics[height=8cm,angle=0]{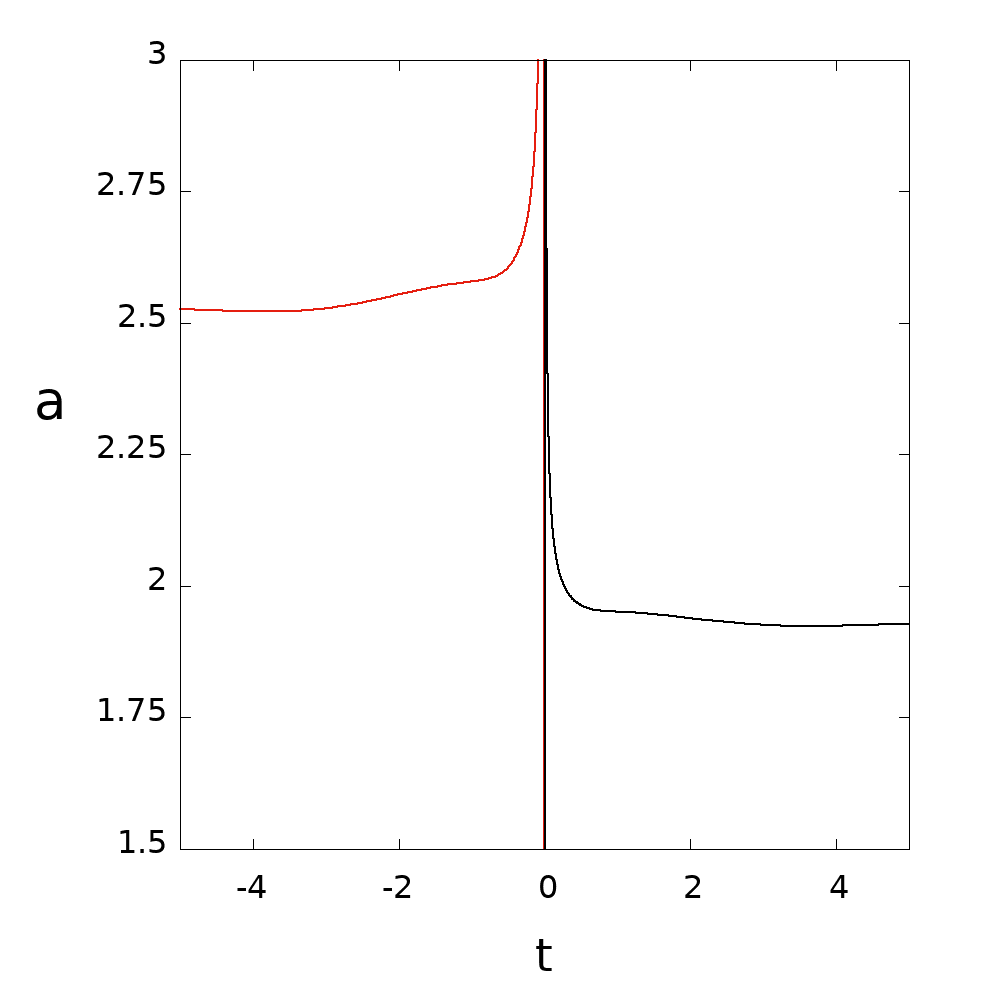}\includegraphics[height=8cm,angle=0]{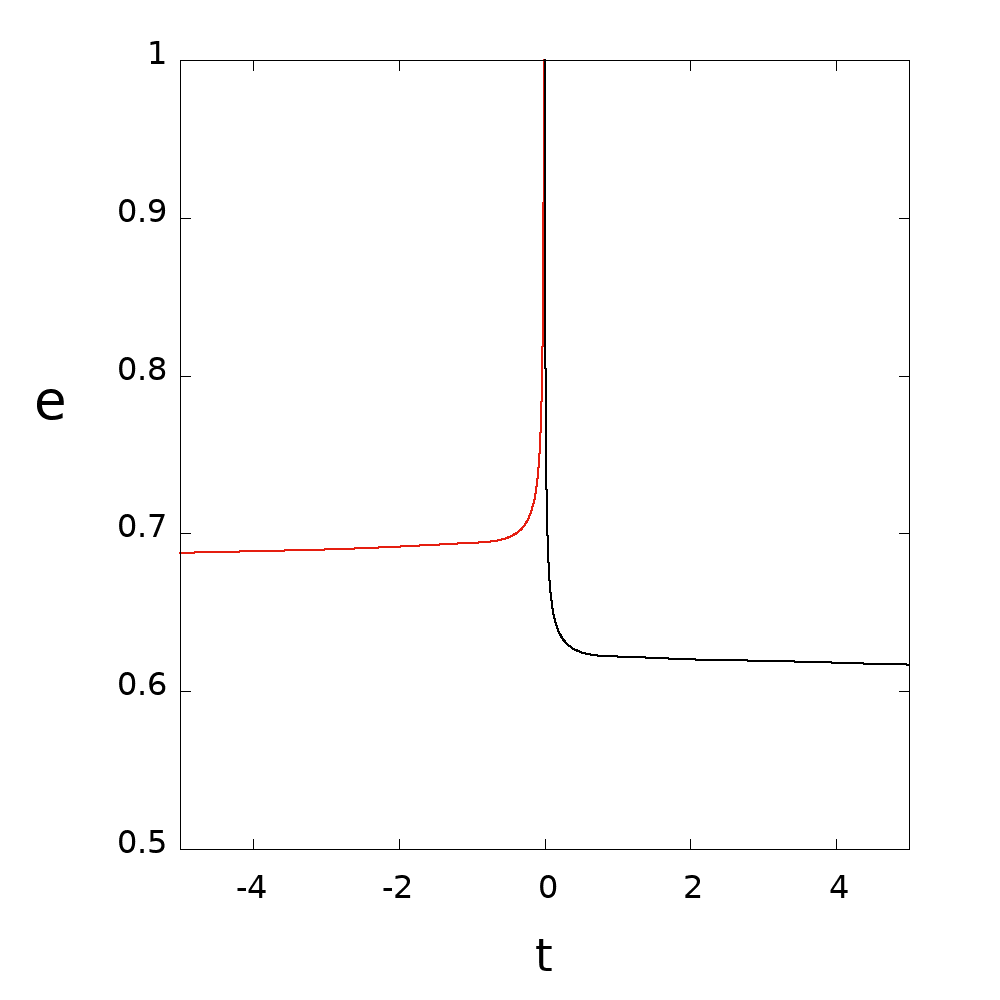}      
  
}
\caption{Representation of the time evolution of the semi-major axis $a$
  and eccentricity $e$ for the orbits represented in Fig. \ref{orbits}: the
  top panels refer to the case $\mu=3\ 10^{-6}$, the bottom panels
  to the case $\mu=10^{-6}$. }
\label{variationsAE}
\end{figure}

The most important parameters of the problem are the mass
parameter $\mu$ and the value $E$ of the Hamiltonian characterizing
the close encounter. We consider two numerical values for the mass parameter $\mu= 10^{-3}$ and $\mu=3.\ 10^{-6}$ which are representative, up to a small correction, of the three-body problems which are defined by considering the couples Sun--Jupiter and Sun--Earth  as primary
bodies respectively.  The method considered in this
paper applies to the fast close encounters, whose  parameter 
$E$ satisfies inequality (\ref{largeE}). Since
the parameter $\lambda=\alpha/2>0$ is a divisor of the Bnf, the values of $\alpha$ close to zero (and in general $\alpha/2 \in (0,1)$)
imply larger coefficients of the monomials of the Bnf
as well as larger errors in the representation of the solutions transiting at a given distance from $P_2$. Therefore, we consider the value of
$E=-1.35$ which is close to the limit value $\alpha=0$
(precisely the small divisor in this case is $\alpha/2 \sim 0.27$);
the results are expected to improve for larger values of  $E$. 

For each value of $\mu$, we choose an initial condition $u(0),U(0)$ well inside the Hill's sphere of the secondary body and we numerically compute its orbit for positive and negative times; this is equivalent to consider orbits having a close encounter with $P_2$. The orbits corresponding to the selected initial conditions are
computed by numerically integrating the Hamilton's equations of the Levi-Civita Hamiltonian using an explicit Runge-Kutta algorithm of order 6, with a very small integration time step of $\tau =10^{-5}$, and
extended floating point numerical precision (the absolute value of
the regularized Hamiltonian remains smaller than $3.5 \ 10^{-16}$).

The numerically computed solutions will be used to check the conservation of the integrals of motions of the first and second Birkhoff Normal Forms, and compared with the solutions provided in parametric form by the Eq. (\ref{focusfocusdynamics}). Two orbits, numerically
computed for positive (black curve) and negative times (red curve), are represented in Fig. \ref{orbits} for the case $\mu=3\ 10^{-6}$ and $\mu=10^{-3}$ respectively. The time variations of the semi-major axis $a$ and eccentricity $e$
are reported in Fig. \ref{variationsAE}: we appreciate that there is a sharp
variation of $a$, $e$ in both cases, associated to the close encounter; 
in the case of the smaller mass value $\mu=3\ 10^{-6}$ the variation is
almost stepwise. 

These orbits are then analyzed using the Birkhoff normalizations described in Sections \ref{sec:firstbnf} and \ref{secondbnf}. The canonical transformations and the normal form Hamiltonian ${\hat H}^N_E$ are computed by implementing with a
computer algebra system the algorithm described in Appendix 1 (Section \ref{chifirstBnf}). The implementation of the Birkhoff transformations requires:
\begin{itemize}
\item[--] the computation of Taylor expansions of functions up to a given
  polynomial order $N$;
\item[--] the computation of the generating functions $\chi_J$ from the Taylor expansion of the Hamiltonian represented with the hyperbolic variables;
\item[--] the computation of the Lie series defined by $\chi_J$ up to
  the degree $N$.
\end{itemize}  
The operations required by these three steps are performed by
several available computer algebra systems. The implementation
can be symbolic, i.e. the coefficients of all the monomials
are given as functions of the parameters $\mu,\alpha$ (and no
floating point approximations are introduced), or numeric, i.e.
for a given value of $\mu,\alpha$ the coefficients of all the monomials are given as floating point numbers. The second choice allows to reach larger
orders $N$. A symbolic implementation has been executed up to $N=20$; a numerical implementation (where $\mu,E$ are treated as floating point numbers) has been executed up to $N=30$, and this is the procedure whose results are reported below. In the numerical implementation the generating functions and the Birkhoff normal forms Hamiltonians are computed as polynomials with floating point coefficients. The canonical transformations, which are the flow at time $\tau=1$ (or
$\tau=-1$ for the inverse) of the generating functions, can be also
represented by polynomials. The direct numerical evaluation
of these polynomials along the flow is computationally expensive;
a less expensive evaluation (which is used in the
computations described below) is obtained by numerically computing
the Hamiltonian flow of the generating functions using an explicit Runge-Kutta algorithm of order 6 with $100$ steps. 
\vskip 0.4 cm
For each orbit:
\begin{itemize}

\item[--]  {we compute the function
       $$
       J^N= q_1p_1+q_2p_2
       $$
       along the numerically computed orbit for different orders of the first Birkhoff normalization ranging from $N=2$ (no normalization) up to $N=30$.
       We also check the conservation of the actions 
       $Q_1P_1$ and $Q_2P_2$ for different normalization orders of the
       second Birkhoff normalization (which are implemented from the
       first Bnf of order $N=30$). Since the real and imaginary parts
       of $Q_1P_1$, $Q_2P_2$ are proportional to $q_1p_1+q_2p_2$ (which
       is already conserved up to order $N=30$ from the first Bnf) or
       $q_1p_2-q_2p_1$, we just need to check the conservation of
$$
W^{N,\tilde N}=\Im(Q_1P_1)
$$
along the numerically computed orbits ($N$ refers to the order of the first Bnf,
$\tilde N$ to  the order of the second Bnf). The relative variations of $J^N,W^{N,\tilde N}$ are represented versus the distance $r=\norm{u}^2$ from the
secondary body, providing us an estimate of the error
which we have  at some distance from $P_2$ using different normalization orders. In fact, the approximation of the solutions of the three-body problem with the solutions of  the Hamilton's equations of Hamiltonian (\ref{hametakappa}) during the transit  inside a sphere centered at $P_2$ of radius $\rho$ is justified if the variations  of the function $J^N,W^{N,\tilde N}$ remain extremely small during the
transit, possibly smaller than the numerical precision.}

\item[--] {for the same initial
  conditions of the numerical integrations we also compute the
  solution provided by the parametric representation of 
  Eq. (\ref{focusfocusdynamics}) (mapped back to the original
  Levi-Civita variables $u,U$). The difference between the parametric and the numerical solution is represented versus the distance $\norm{u}^2$ for different normalization orders.}

\end{itemize}

\begin{figure}[!]
\center{
\includegraphics[height=8cm,angle=0]{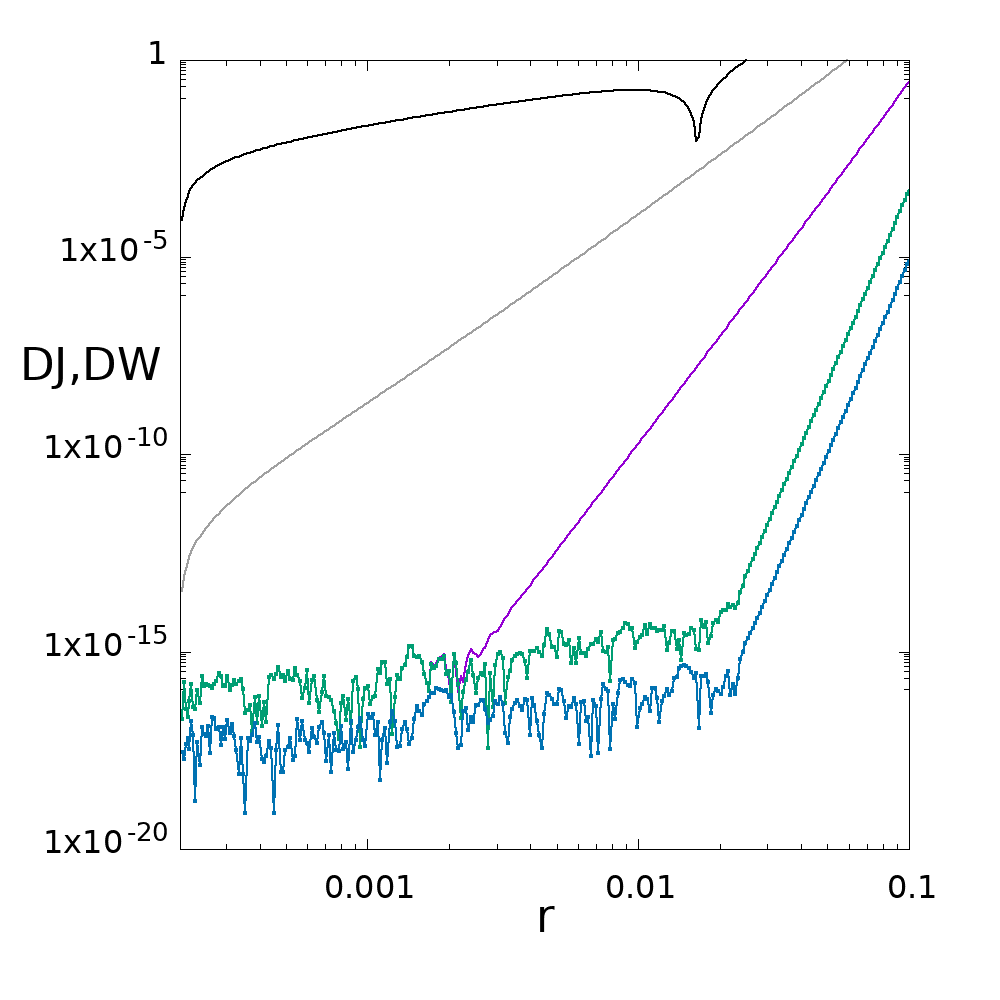}\includegraphics[height=8cm,angle=0]{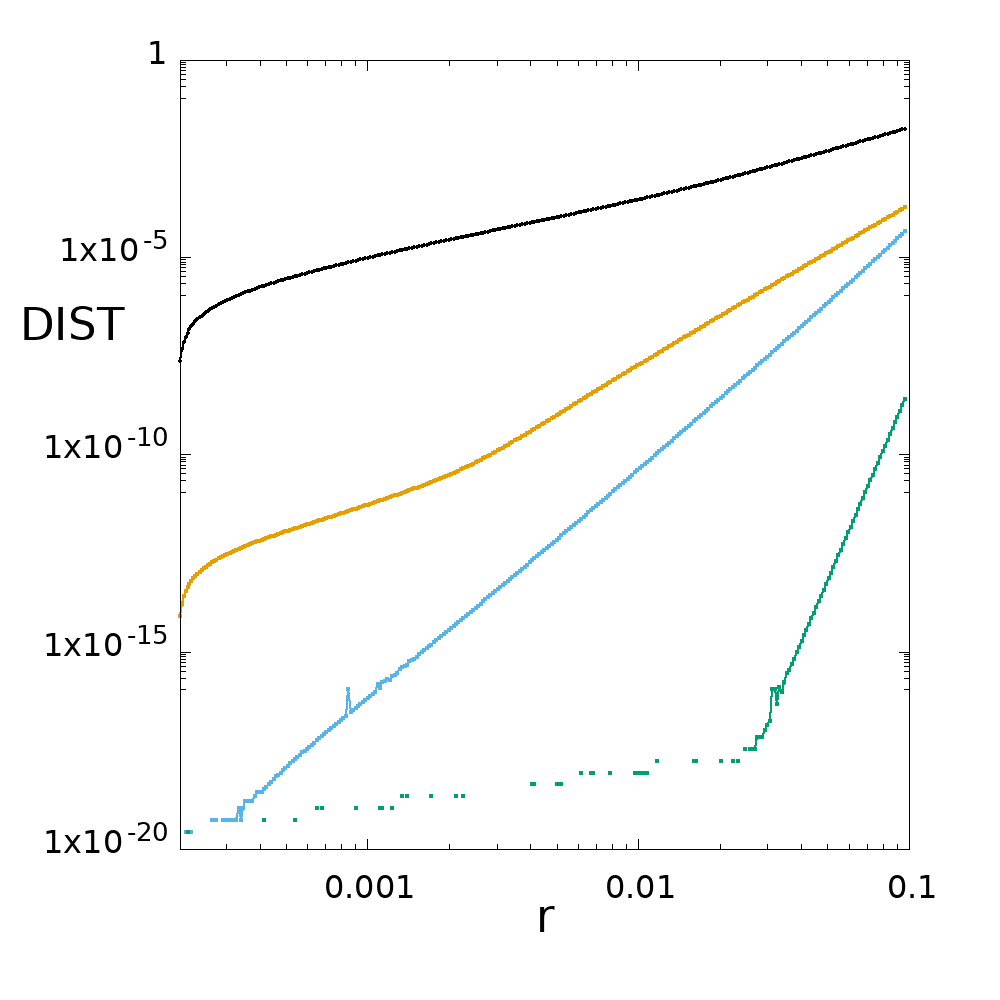}
}

\caption{Left panel: Relative variation ${\rm DJ}$ (defined in Eq. (\ref{DJ})) computed for the same  orbit of Fig.  \ref{orbits}, top panels
  (case $\mu =3\ 10^{-6}$) represented versus the distance
  $r:=\norm{u}^2$ from $P_2$ for $N=2$ (black) and the improvements
  provided by the Birkhoff normalizations with 
 $N=8$ (gray),  $N=16$,
  (violet),  $N=30$ (green).  The blue line represents instead
  the relative variation ${\rm DW}$ (defined in Eq. (\ref{DW})) computed for $N=30,\tilde N=12$.
  Right panel:  variation of ${\rm DIST}$ (defined in Eq. (\ref{dist})) 
 represented versus the distance
$r:=\norm{u}^2$ for $N,\tilde N=2$ (black) and 
the improvements provided by the Birkhoff normalizations with 
$N,\tilde N =4$ (gold),
 $N,\tilde N=10$ (light blue),  $N=30,\tilde N=14$ (green).  
}
\label{J-evolutionearth}
\end{figure}

\begin{figure}[!]
\center{

\includegraphics[height=8cm,angle=0]{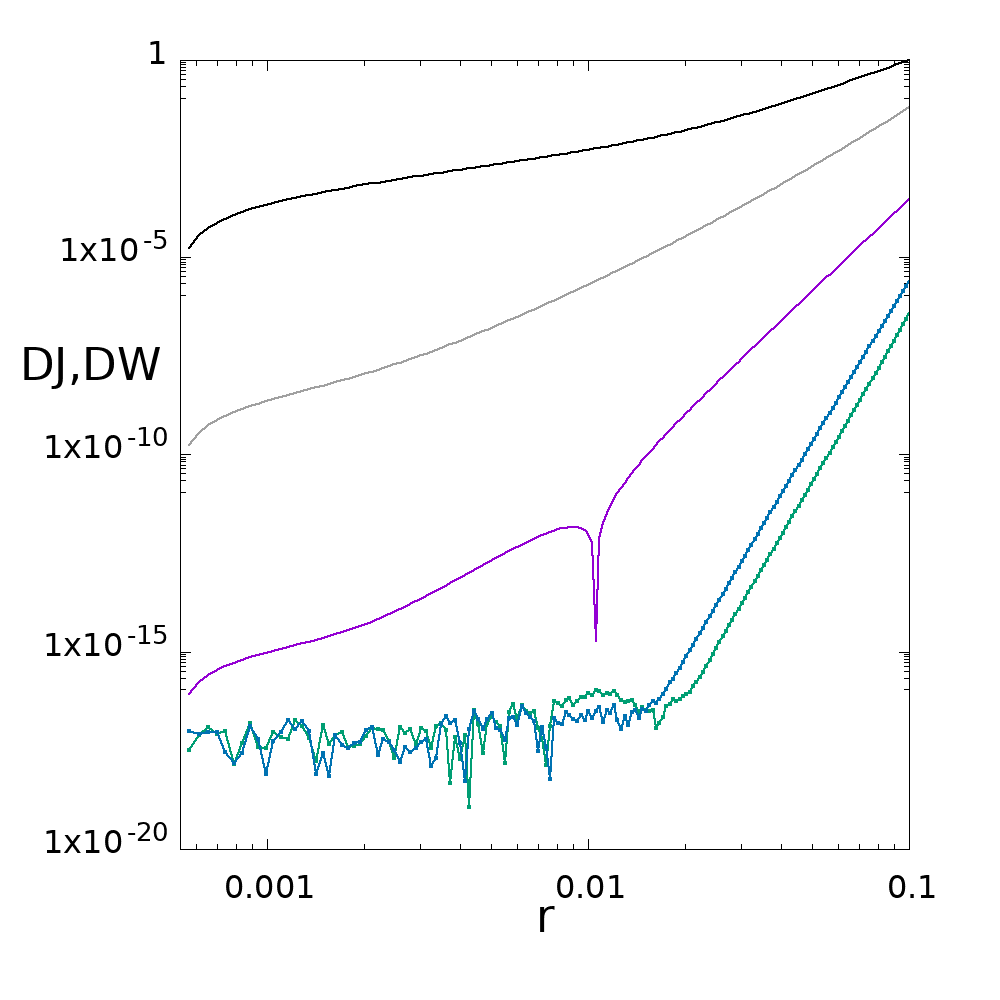}\includegraphics[height=8cm,angle=0]{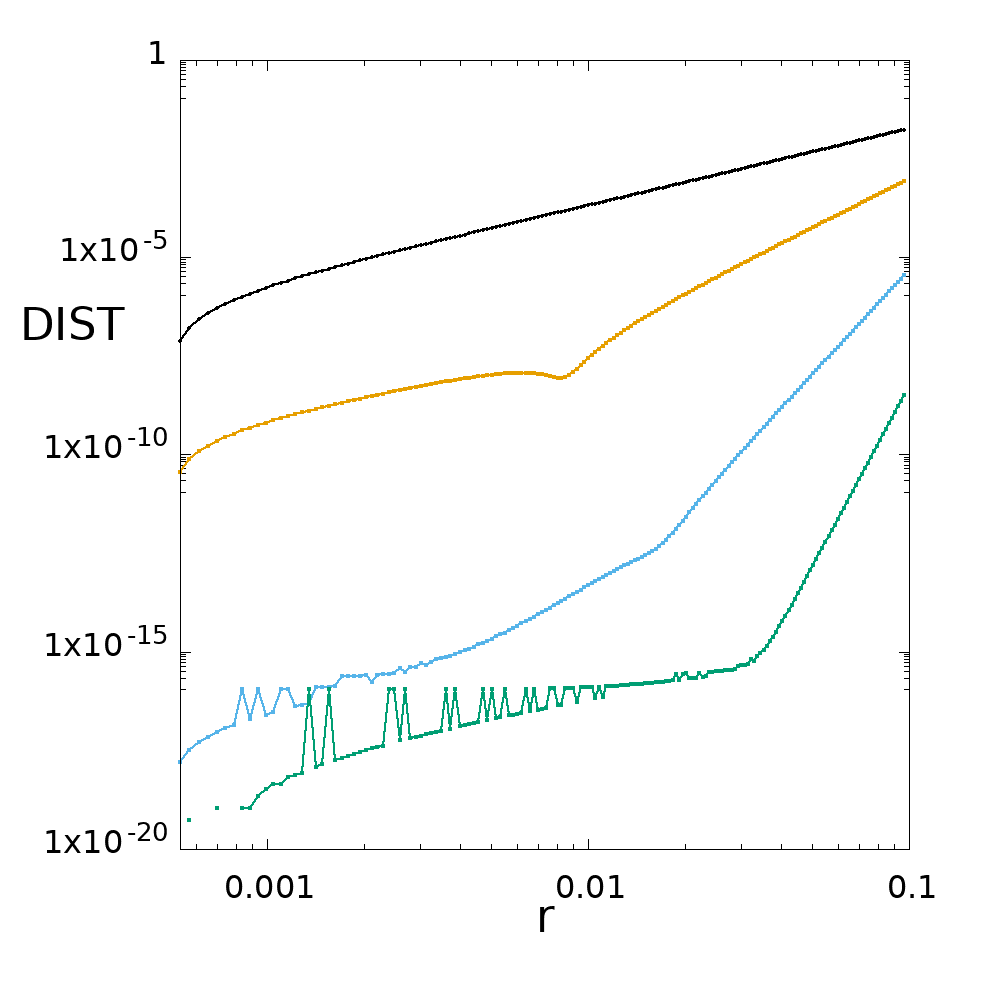}
}
\caption{Left panel: Relative variation $DJ$  (defined in Eq. (\ref{DJ}))
computed for the same  orbit of Fig.  \ref{orbits}, bottom panels
  (case $\mu =10^{-3}$) represented versus the distance
  $r:=\norm{u}^2$ for $N=2$ (black) and the improvements
  provided by the Birkhoff normalizations with 
 $N=8$ (gray),  $N=16$,
  (violet),  $N=30$ (green). 
The blue line represents instead
  the relative variation ${\rm DW}$ (defined in Eq. (\ref{DW})) computed for $N=30,\tilde N=16$.
  Right panel:  variation of ${\rm DIST}$ (defined in Eq. (\ref{dist})) represented versus the distance
  $r:=\norm{u}^2$ for $N=\tilde N=2$ (black),
 and the improvements provided by the Birkhoff normalizations with 
$N,\tilde N =4$ (gold),
 $N,\tilde N=14$ (light blue),  $N=30,\tilde N=14$ (green).  
}
\label{J-evolutionjupiter}
\end{figure}

During the fast close encounters, the orbits have a sharp change of
the semi-major axis and eccentricity (see Fig. \ref{variationsAE}).
In the left panels of  Fig.s \ref{J-evolutionearth} and \ref{J-evolutionjupiter} we represent the relative variations:
\begin{equation}
  {\rm DJ}={\norm{J^N(q(t),p(t))-J^N(q(0),q(0))}\over \norm{
      J^N(q(0),q(0)) }}
  \label{DJ}
\end{equation}
for different orders of the first Birkhoff normalization ranging from $N=2$ (no normalization) up to $N=30$, computed 
during the close encounter occurring from initial time until 
$P$ reaches a distance $r=0.1$ from $P_2$, for the orbit of the
case $\mu=3\ 10^{-6}$  (Fig. \ref{J-evolutionearth}) and 
$\mu=10^{-3}$ (Fig. \ref{J-evolutionjupiter}). The relative
  variations are indeed extremely small for $N=30$
  ($\sim 10^{-15}$ ) up to $r \sim 0.02$ for both orbits,
  with a possible improvement which still could be obtained by considering  larger values of $N$.
  In the same panels we also report the relative variation of $W^{N,\tilde N}$,
computed after the $N=30$ normalizations of the first Bnf:
\begin{equation}
  {\rm DW}={\norm{W^{N,\tilde N}(q(t),p(t))-W^{N,\tilde N}(q(0),q(0))}\over \norm{
      W^{N,\tilde N}(q(0),q(0)) }} ,
  \label{DW}
\end{equation}
which is very small  up to $r \sim 0.02$ already for $\tilde N=10$ (Fig. \ref{J-evolutionearth}) and $\tilde N=16$ (Fig. \ref{J-evolutionjupiter}) respectively.  

In the left panels of (Fig. \ref{J-evolutionearth}) and (Fig. \ref{J-evolutionjupiter}) we represent the distance:
\begin{equation}
{\rm DIST}= \Norm{ (x^N-x^P,y^N-y^P)}
\label{dist}
\end{equation}
in Cartesian variables $(x,y)$  between the solutions $(x^N,y^N)$ 
computed by numerically integrating the Hamilton's equations
of the Levi Civita Hamiltonian and the solutions $(x^P,y^P)$ provided by the parametric representation  (\ref{focusfocusdynamics}): the distances are smaller than $10^{-16}$ up to $r=0.02$.

 \section{Appendix 1: definition of the Birkhoff normal forms for
 2-degrees of freedom Hamiltonians}\label{chifirstBnf}

 Consider the 2-degrees of freedom Hamiltonian system
 of the form:
\begin{equation}
H(q,p)= h_2(q,p)+h_4(q,p)+h_6(q,p)....
\label{hamH}
\end{equation}
where:
$$
h_2(q,p)=\lambda_1 q_1p_1 + \lambda_2 q_2p_2  ,
$$
and the functions $h_{2j}(q,p)$ are polynomials of degree $2j$.

For any even number $N\geq 4$, we apply 
iteratively a finite sequence of canonical transformations,
called Birkhoff normalizations, whose composition is a
close to the identity
canonical transformation defined in a neighbourhood of
$(q,p)=(0,0,0,0)$ (which is a fixed point of the transformation), 
conjugating $H$ to
\begin{equation}
H^N(q,p)=  h_2(q,p)+{\hat h}_4(q,p)+\ldots
+ {\hat h}_N(q,p)+{\cal R}_{N+2}(q,p)
\label{firstBNFgeneric}
\end{equation}
where ${\cal R}_{N+2}$ is a series in $q,p$ starting with
terms of degree at least $N+2$ and the ${\hat h}_j$ are polynomials
of order $j$ in $q,p$ (only
the polynomials with $j$ even number appear in the expansion)
containing monomials:
\begin{equation}
c^{j,N}_{m,n}q_1^{m_1}q_2^{m_2}p_1^{n_1}p_2^{n_2}
\label{monomialgeneric}
\end{equation}
with $m=(m_1,m_2),n=(n_1,n_2)$ satisfying:
\begin{equation}
  \lambda_1 (n_1-m_1)+\lambda_2 (n_2-m_2)=0 .
   \label{resmonomialsgeneric}
\end{equation}
The Hamiltonian which is obtained by neglecting the remainder ${\cal R}_{N+2}(q,p)$ in ${ H}^N(q,p)$:
$$
{\hat H}^N(q,p)= h_2(q,p)+{\hat h}_4(q,p)+\ldots
+ {\hat h}_N(q,p)  ,
$$
will be called Birkhoff normal form of $H$ of order $N$. The  Birkhoff transformations and the transformed Hamiltonians can be computed explicitly with a
computer algebra system, by representing them with the Lie series method
(for a modern presentation of the Lie series method we refer to
\cite{giorgillibook}), following the procedure described below. 

The Birkhoff normalization of Hamiltonian (\ref{hamH}) is obtained, for
any $N\geq 4$, with a canonical transformation ${\cal C}_N$ defined as the composition of a sequence of canonical Birkhoff transformations:
\begin{equation}
 (q,p):= (q^{(2)},p^{(2)}) \mapsto  (q^{(4)},p^{(4)}) \mapsto \ldots \mapsto
  (q^{(N)},p^{(N)})
\end{equation}
where each elementary transformation:
\begin{equation}
(q^{(J-2)},p^{(J-2)}) = \Phi^1_{\chi_{_J}}(q^{(J)},p^{(J)}))
\label{elementary}
\end{equation}  
is the Hamiltonian flow $\Phi^1_{\chi_{_J}}$ at time $1$ of an Hamiltonian
$\chi_{_J}$, which is polynomial of degree $J$. The elementary transformations
(\ref{elementary}) can be computed explicitly as the Lie series:
\begin{equation}\label{eq:theCchin}
 \zeta = \Big [ e^{\,L_{\chi_{_J}}} \zeta \Big ](q^{(J)},p^{(J)}) := \zeta' +
 \{\zeta, \chi_{_J} \}(q^{(J)},p^{(J)})+ {1\over 2} \{ \{\zeta,
 \chi_{J} \}, \chi_{_J} \}(q^{(J)},p^{(J)})+\ldots ~,
\end{equation}
where $L_{\chi_{_J}} := \{\cdot, \chi_{_J} \}$, and $\zeta, \zeta'$ denote any couple of variables
${{q}}^{(J-2)},{{q}}^{(J)}$ or 
${{p}}^{(J-2)},{{p}}^{(J)}$. The Hamiltonian $H^{(2)}:=H$ is conjugate
to the sequence of Hamiltonians $H^{(4)},\ldots ,H^{(J)}$ which can 
be computed as Lie series as well:
\begin{equation}\label{eq:newham}
  {H}^{(J)} = e^{L_{\chi_{_J}}}\, {H}^{(J-2)} ,
\end{equation}
and the iteration ends for $J=N$. The aim is to define the generating functions
$\chi_{_J}$ such that, at any step $J$, the intermediate
normalized Hamiltonians:
\begin{equation}\label{eq:intermediateminormf}
{H}^{(J)} :=  h_2+\sum_{j=2}^{J\over 2} {H}^{(J)}_{2j}+
\sum_{j\geq (J+2)/2} {\tilde H}^{(J)}_{2j} 
\end{equation}
have the property that $H^{(J)}_{2j}$ and  ${\tilde H}^{(J)}_{2j}$ are polynomials of degree $2j$, and moreover $H^{(J)}_{2j}$
contain only monomials satisfying:
\begin{equation}
  \lambda_1 (n_1-m_1)+\lambda_2 (n_2-m_2)=0 .
   \label{resmonomials}
\end{equation}
We suppose that
the Hamiltonian has been normalized up to given $J\geq 4$, and we
define the subsequent normalization with a  generating function $\chi_{_{J+2}}$
which is polynomial of degree $J+2$. Therefore, from the representation
(\ref{eq:intermediateminormf}), we have:
$$
{H}^{(J+2)} =  h_2+\sum_{j=2}^{J\over 2} {H}^{(J)}_{2j}
+(\{ h_2, \chi_{_{J+2}}\} +  {\tilde H}^{(J)}_{J+2} ) + {\cal O}(J+4)  ,
$$
whose term of degree $J+2$ is $\{ h_2, \chi_{_{J+2}}\} +  {\tilde H}^{(J)}_{J+2}$. 
By denoting with $a^{(J)}_{m_1,m_2,n_1,n_2}q_1^{m_1}q_2^{m_2}p_1^{n_1}p_2^{n_2}$ the monomials of ${\tilde H}^{(J)}_{J+2}$, and
defining the generating function ${\chi_{_{J+2}}}$ by:
  \begin{equation}\label{eq:genfunc-a}
  \chi_{_{J+2}} = \hspace{-0.6cm}
  \sum_{\substack{(m_1,m_2,n_1,n_2)\in {\cal L}_{J+2}}}
  \hspace{-0.3cm}
    \frac{a^{(J)}_{m_1,m_2,n_1,n_2}}{
      [\lambda_1 (m_1-n_1)+\lambda_2(m_2-n_2)]} \,
    {q}_1^{m_1} {q}_2^{m_2} {p}_1^{n_1}
    {p}_2^{n_2}
\end{equation}
where:
\begin{displaymath}
{\cal L}_{_J}=\left\{ (m_1,m_2,n_1,n_2)\in {\Bbb N}^4:  \sum_{j=1}^2
(n_j+m_j)=J,\ {\rm and}\ \ \lambda_1 (m_1-n_1)+\lambda_2(m_2-n_2)  \ne 0 \right\}~,
\end{displaymath}    
we have
\begin{equation}
  \{ h_2, \chi_{_{J+2}}\} +  {\tilde H}^{(J)}_{J+2} =  \Pi {\tilde H}^{(J)}_{J+2}
  :=  \sum_{m,n: \norm{m}+\norm{n}=J+2, 
\lambda_1 (n_1-m_1)+\lambda_2(n_2-m_2)=0  }a^{(J)}_{m,n}q^mp^n  .
\label{feq1}
\end{equation}
Therefore, ${H}^{(J+2)}$ is normalized up to degree $J+2$. 
Since the generating function $\chi_{_J}$ is polynomial of order
larger than $4$, its flow at time 1 is close to the identity and
has the origin $(q,p)=(0,0,0,0)$ as a fixed point.

\section{Appendix 2: the Birkhoff normalization for $N=6$}\label{appendix2}

The saddle-saddle Birkhoff normalization of order $N=6$ is obtained using the
generating functions:
{\small
$$
\chi_4 = \frac{{p_1}^2 {p_2} {q_1}}{8 \alpha ^2}-\frac{{p_1}^3 {q_2}}{8 \alpha
   ^2}-\frac{{p_1} {p_2}^2 {q_2}}{8 \alpha ^2}+\frac{{p_1} {q_1}^2 {q_2}}{8
   \alpha ^2}+\frac{{p_1} {q_2}^3}{8 \alpha ^2}+\frac{{p_2}^3 {q_1}}{8 \alpha
  ^2}-\frac{{p_2} {q_1}^3}{8 \alpha ^2}-\frac{{p_2} {q_1} {q_2}^2}{8 \alpha ^2}
$$
}
{\small
\begin{dmath*}
\chi_6 = \frac{{p_1}^2 {p_2}^2 {q_1}^2 (9 {\mu }-11)}{32 \alpha ^4}-\frac{3 {p_1}^3
   {p_2}^2 {q_1} ({\mu }-1)}{32 \alpha ^4}+\frac{{p_1}^2 {p_2}^2 {q_2}^2 (9
   {\mu }-11)}{32 \alpha ^4}-\frac{3 {p_1}^2 {p_2}^3 {q_2} ({\mu }-1)}{32
   \alpha ^4}+\frac{{p_1}^4 {p_2}^2 ({\mu }-1)}{64 \alpha ^4}+\frac{{p_1}^2
   {p_2}^4 ({\mu }-1)}{64 \alpha ^4}+\frac{{p_1}^3 {p_2} {q_1} {q_2} (3
   {\mu }-2)}{8 \alpha ^4}-\frac{3 {p_1}^4 {p_2} {q_2} ({\mu }-1)}{64 \alpha
   ^4}+\frac{{p_1}^2 {q_1}^2 {q_2}^2 (11-9 {\mu })}{32 \alpha ^4}-\frac{15
   {p_1}^4 {q_1}^2 ({\mu }-1)}{64 \alpha ^4}+\frac{15 {p_1}^2 {q_1}^4
   ({\mu }-1)}{64 \alpha ^4}+\frac{3 {p_1}^5 {q_1} ({\mu }-1)}{64 \alpha
   ^4}+\frac{{p_1}^4 {q_2}^2 (3 {\mu }-7)}{64 \alpha ^4}+\frac{{p_1}^2 {q_2}^4
   (7-3 {\mu })}{64 \alpha ^4}-\frac{{p_1}^6 ({\mu }-1)}{192 \alpha
   ^4}+\frac{{p_1} {p_2}^3 {q_1} {q_2} (3 {\mu }-2)}{8 \alpha ^4}-\frac{3
   {p_1} {p_2}^4 {q_1} ({\mu }-1)}{64 \alpha ^4}+\frac{{p_1} {p_2}
   {q_1}^3 {q_2} (2-3 {\mu })}{8 \alpha ^4}+\frac{{p_1} {p_2} {q_1}
   {q_2}^3 (2-3 {\mu })}{8 \alpha ^4}+\frac{3 {p_1} {q_1}^3 {q_2}^2
   ({\mu }-1)}{32 \alpha ^4}-\frac{3 {p_1} {q_1}^5 ({\mu }-1)}{64 \alpha
   ^4}+\frac{3 {p_1} {q_1} {q_2}^4 ({\mu }-1)}{64 \alpha ^4}+\frac{{p_2}^2
   {q_1}^2 {q_2}^2 (11-9 {\mu })}{32 \alpha ^4}+\frac{{p_2}^2 {q_1}^4 (7-3
   {\mu })}{64 \alpha ^4}+\frac{{p_2}^4 {q_1}^2 (3 {\mu }-7)}{64 \alpha
   ^4}+\frac{15 {p_2}^2 {q_2}^4 ({\mu }-1)}{64 \alpha ^4}-\frac{15 {p_2}^4
   {q_2}^2 ({\mu }-1)}{64 \alpha ^4}+\frac{3 {p_2}^5 {q_2} ({\mu }-1)}{64
   \alpha ^4}-\frac{{p_2}^6 ({\mu }-1)}{192 \alpha ^4}+\frac{3 {p_2} {q_1}^2
   {q_2}^3 ({\mu }-1)}{32 \alpha ^4}+\frac{3 {p_2} {q_1}^4 {q_2}
   ({\mu }-1)}{64 \alpha ^4}-\frac{3 {p_2} {q_2}^5 ({\mu }-1)}{64 \alpha
   ^4}-\frac{{q_1}^2 {q_2}^4 ({\mu }-1)}{64 \alpha ^4}-\frac{{q_1}^4 {q_2}^2
   ({\mu }-1)}{64 \alpha ^4}+\frac{{q_1}^6 ({\mu }-1)}{192 \alpha
  ^4}+\frac{{q_2}^6 ({\mu }-1)}{192 \alpha ^4}
\end{dmath*}}
\noindent
The focus-focus Birkhoff normalization of order $N=6$ is obtained using the
generating functions:
{\small
$$
\chi_4=\frac{15 i {P_2}^2 {Q_1}^2 (\mu-1)}{32 \alpha ^2}-\frac{15 i {P_1}^2 {Q_2}^2
  (\mu-1)}{32 \alpha ^2}
$$
\begin{dmath*}
  \chi_6= \frac{15 i {P_1}^2 {P_2} {Q_1}^3 (\mu-1)}{128 \alpha ^3}-\frac{75 i {P_1}^2
   {P_2} {Q_1} {Q_2}^2 (\mu-1)}{128 \alpha ^3}-\frac{45 {P_1}^2 {P_2}
   {Q_2}^3 \left(\mu^2-6 \mu+5\right)}{256 \alpha ^4}-\frac{15 i {P_1}^3
   {Q_1}^2 {Q_2} (\mu-1)}{128 \alpha ^3}+\frac{45 {P_1}^3 {Q_1} {Q_2}^2
   \left(\mu^2-6 \mu+5\right)}{256 \alpha ^4}-\frac{175 i {P_1}^3 {Q_2}^3
   (\mu-1)}{384 \alpha ^3}+\frac{75 i {P_1} {P_2}^2 {Q_1}^2 {Q_2}
   (\mu-1)}{128 \alpha ^3}-\frac{45 {P_1} {P_2}^2 {Q_1}^3 \left(\mu^2-6
   \mu+5\right)}{256 \alpha ^4}-\frac{15 i {P_1} {P_2}^2 {Q_2}^3
   (\mu-1)}{128 \alpha ^3}+\frac{45 {P_2}^3 {Q_1}^2 {Q_2} \left(\mu^2-6
   \mu+5\right)}{256 \alpha ^4}+\frac{175 i {P_2}^3 {Q_1}^3 (\mu-1)}{384
    \alpha ^3}+\frac{15 i {P_2}^3 {Q_1} {Q_2}^2 (\mu-1)}{128 \alpha ^3}
\end{dmath*}
}

\section{Conclusions and perspectives}\label{sec:Conclusions}

The method presented in this paper allows to compute the
close encounters with the secondary body of the planar circular
restricted three--body problem using integrable dynamics
approximating the regularized Hamiltonian at any order $N$ of
its Birkhoff approximations. The numerical implementations of the
method show that it indeed can be applied to problems whose parameters are compatible with important Solar System three-body problems,
with errors on the predicted orbits which can be reduced below
the round-off approximation. Therefore, it provides an Hamiltonian 
integrator of the close encounters as a single step. 

We have therefore the opportunity to compare the output of
numerical integrations with these Hamiltonian dynamics to confirm
their correctness when they agree below a precision threshold. 
In fact, there is no way to state the correctness of the numerical integration
of a close encounter, a part from the necessary conservation of the Hamiltonian
(which implies the approximate conservation of the Tisserand parameter
before and after the close encounter). 

In future researches, applications of the method for a full range of parameters $\mu,E$ which are relevant for Solar system applications will be investigated, as well as possible extensions to models which are more representative of the dynamics of a realistic model of the Solar System.

\vskip 0.4 cm

\noindent
    {\bf Acknowledgments.} The author thanks prof.s  E. Detomi and A. Lucchini
    for presenting to me the sufficient condition for the
    factorization of
    polynomials of four variables. He also acknowledges the project
    MIUR-PRIN 20178CJA2B ``New frontiers of
    Celestial Mechanics: theory and applications".


\begin{thebibliography}{30}


 \bibitem{arenstorf}
  Arenstorf R.F., Periodic solutions of the restricted three-body problem representing analytic
continuations of Keplerian elliptic motions, Amer. J. Math., 85, pp. 27-35, 1963.

\bibitem{bolotinmkkay2000}
Bolotin, S.V., MacKay, R.S.: Periodic and chaotic trajectories of the second species for the n-centre problem.
Celest. Mech. Dyn. Astron. 77, 49–75, 2000. 

\bibitem{bolotin2006}
Bolotin S., Shadowing chains of collision orbits. Discr. Conts. Dyn. Syst. 14, 235–260, 2006.



\bibitem{capinskietal2023}
  Capi\'nski M.J., Kepley S., Mireles James, J.D.,
Computer assisted proofs for transverse collision and near collision orbits in the restricted three body problem, Journal of Differential Equations, Vol, 366, 132-191, 2023.

 \bibitem{CG21}
	Cardin F. and Guzzo M. ``Integrability of close encounters in the spatial restricted three-body problem''. Accepted for publication by \emph{Comm. Cont. Math.} 2021.


\bibitem{CCP16}
Ceccaroni, M., Celletti, A. and Pucacco G.,
Halo orbits around the collinear points of the restricted three-body problem.
Physica D, Volume 317, 1, p. 28-42, 2016.

  

\bibitem{chencinerllibre}        
Chenciner A., Llibre J., A note on the existence of invariant punctured tori in the planar circular restricted three-body problem, Ergod. Theory Dyn. Syst. (Charles Conley Memorial Issue) 8, 63–72, 1988. 

\bibitem{clairaut}
Clairaut A.-C., Th\`eorie du mouvement des com\`etes, Paris, 1760. 

\bibitem{fejoz2002}
F\'ejoz J., Quasiperiodic motions in the planar three-body problem, J. Differ. Equ. 183, 2, 303–341, 2002.


\bibitem{FNS}
{Font J., Nunes A., Sim\'o C., Consecutive quasi-collisions in the planar circular RTBP, Nonlinearity, 15, 115, 2002.}

\bibitem{FNS2009}
Font J., Nunes A., Sim\'o C., A numerical study of the orbits of second species of the planar circular RTBP, Celest.
Mech. Dyn. Astron. 103, 2, 143–162, 2009.

\bibitem{giorgillibook}
  Giorgilli A., Notes on Hamiltonian Dynamical Systems.
  Cambridge University Press, 2022.   

\bibitem{GJMS}
  G\'omez G., Jorba \`A., Masdemont J.,  Sim\'o C., Dynamics and Mission Design Near Libration Point Orbits, Vol. 3: Advanced Methods for Collinear Points, World Scientific, Singapore, 2000.

\bibitem{Jacobson}
  Jacobson N., Basic Algebra 1, W. H. Freeman and Co., San Francisco, CA, 1974,
 second edition 1985.
  

\bibitem{JM99}
Jorba A., Masdemont J., Dynamics in the center manifold of the restricted three-body problem, Physica D 132, 189-213, 1999.

  
\bibitem{GKZ}
{Guardia M., Kaloshin V., Zhang J., Asymptotic Density of Collision Orbits in the Restricted Circular Planar 3 Body Problem, Archive for Rational Mechanics and 
Analysis, vol. 233, Issue 2, pp 799-836, 2019.}

\bibitem{halley}
{Halley E., A Synopsis of the Astronomy of Comets, 1705.}

\bibitem{henon}
H\'enon, M.R.: Generating Families in the Restricted Three Body Problem, Lect. Notes in Phys. Monographs,
52, Springer, Berlin, 1997. 

   \bibitem{henrard}
{Henrard J. ``On Poincar\'e's second species solutions''. In \emph{Cel. Mech.},
        vol 21, 83-97, 1980. }

	\bibitem{LC1906} 
	  Levi-Civita T. ``Sur la régularisation qualitative du probléme restreint des trois corps''. In: \emph{Acta Math.} 30 (1906), pp. 305-327.

        \bibitem{Laplace}
          Laplace P.S., Trait\'e de M\'ecanique C\'eleste, T. IV, Livre IX,
          Chapitre II, Paris, 1880. 
          
\bibitem{Leverrier}
Le Verrier U.J.,	
Th\'eorie de la comete p\'eriodique de 1770, 
Annales de l'Observatoire imperial de Paris, Memoires, t. 3. Paris: 
Mallet-Bachelier, p. 203-270, 1-12, 1857.




\bibitem{marconiederman}
  Marco J.-P., Niederman L., Sur la construction des solutions de seconde espèce dans le probl\`eme plan restreint
  des trois corps, Ann. Inst. H. Poincaré Phys. Théor. 62, 211–249, 1995.


\bibitem{masdemont05}
{ Masdemont J.J., High Order Expansions of Invariant Manifolds of
Libration Point Orbits with Applications to Mission Design, Dynamical
Systems: An International Journal, vol. 20, no. 1, pp. 59-113, 2005.}
  

\bibitem{Perko}
  Perko L.M., Periodic Orbits in the Restricted Three-Body Problem: Existence and Asymptotic Approximation, Siam J. Appl. Math. 41, 200-237, 1974.
  

\bibitem{PG20}
Paez R.I., Guzzo M.,
A study of temporary captures and collisions in the Circular Restricted
Three-Body Problem with normalizations of the Levi-Civita Hamiltonian,
Int. J. of Non-Lin. Mech., 120, 103417, 2020. 
  
     \bibitem{PG21}
       Paez R.I. and Guzzo M. Transits close to the Lagrangian solutions $L_1$, $L_2$ in the elliptic restricted three-body problem, \emph{Nonlinearity},  34 (2021), 6417-6449.

\bibitem{PG22}
Paez R.I., Guzzo M., On the semi-analytical construction of halo orbits
and halo tubes in the elliptic restricted three-body problem,
Physica D: Nonlinear Phenomena 439, 133402, 2022.
          
\bibitem{petersonetal2023}          
 Peterson L.T., Rosales J.J., Scheeres D.J., The Vicinity of Earth-Moon L1 and L2 in the Hill Restricted 4-Body Problem, Physica D: Nonlinear Phenomena, vol. 45, 133889, 2023.

          
\bibitem{pucacco19}
Pucacco G.,
Structure of the centre manifold of the $L_1,L_2$ collinear 
libration points in the restricted three-body problem,
Cel. Mech. and Dyn. Astr., 131, article number 44 (2019).

\bibitem{Rosalesetal2023}
Rosales J.J., Jorba \`A., Jorba-Cusc\'o M., Invariant manifolds near L1 and L2 in the quasi-bicicurlar problem, Celestial Mechanics and Dynamical Astronomy 135, 15, 2023.

\bibitem{Simo99}
Sim\'o C., 
{ Dynamical systems methods for space missions on a vicinity of 
collinear libration points}, in Sim\'o, C., editor, Hamiltonian Systems
with Three or More Degrees of Freedom (S'Agar\'o, 1995), volume 533 of
NATO Adv. Sci. Inst. Ser. C Math. Phys. Sci., pp. 223-241, Dordrecht.
Kluwer Acad. Publ., 1999.

\bibitem{tisserand}
Tisserand F.F., 
Sur la th\'eorie de la capture de com\`etes p\'eriodiques, Bulletin Astronomique, vol. 6, p. 241-257, 1889.
 

\bibitem{zhao2015}
  Zhao L., Quasi-periodic almost-collision orbits in the spatial three-body problem, Commun. Pure Appl. Math.
68, 12, 2144–2176, 2015. 


\end{thebibliography}
\end{document}